\documentclass[twocolumn]{aastex63}

\usepackage{amsmath}
\usepackage{amssymb}
\usepackage{booktabs}
\usepackage[activate={true,nocompatibility},final,tracking=true,kerning=true,spacing=true,factor=1000,stretch=10,shrink=10]{microtype}
\graphicspath{{./}{figures/}}

\newcommand{\pzcode}[1]{\texttt{#1}}
\newcommand{\bpz}{\pzcode{BPZ}}

\begin{document}


\title{Learning Spectral Templates for Photometric Redshift Estimation \\ from Broadband Photometry}

\correspondingauthor{John Franklin Crenshaw}
\email{jfc20@uw.edu}

\author[0000-0002-2495-3514]{John Franklin Crenshaw}
\affiliation{Department of Physics, University of Washington, Box 351560, Seattle,
WA 98195}

\author[0000-0001-5576-8189]{Andrew J. Connolly}
\affiliation{DIRAC Institute, Department of Astronomy, University of Washington, Box 351580, Seattle, WA 98195}
\affiliation{eScience Institute, University of Washington, Box 351570, Seattle, WA 98195}

\begin{abstract}
    Estimating redshifts from broadband photometry is often limited by how accurately we can map the colors of galaxies to an underlying spectral template. 
    Current techniques utilize  spectrophotometric samples of galaxies or spectra derived from spectral synthesis models. 
    Both of these approaches have their limitations, either the sample sizes are small and often not representative of the diversity of galaxy colors or the model colors can be biased (often as a function of wavelength) which introduces systematics in the derived redshifts. 
    In this paper we learn the underlying spectral energy distributions from an ensemble of  $\sim$100K galaxies with measured redshifts and colors. 
    We show that we are able to reconstruct emission and absorption lines at a significantly higher resolution than the broadband filters used to measure the photometry for a sample of 20 spectral templates. 
    We find that our training algorithm reduces the fraction of outliers in the derived photometric redshifts by up to 28\%, bias up to 91\%, and scatter up to 25\%, when compared to estimates using a standard set of spectral templates. 
    We discuss the current limitations of this approach and its applicability for recovering the underlying properties of galaxies.
    Our derived templates and the code used to produce these results are publicly available in a dedicated Github repository: \url{https://github.com/dirac-institute/photoz_template_learning}.
    \\
    \medskip
\end{abstract}


\section{Introduction}
    Studies of galaxy evolution, galaxy clusters, large-scale structure, weak lensing, etc all rely on the determination of galaxy redshift.
    Spectroscopic surveys of galaxies can provide very accurate redshifts by measuring the shifted wavelengths of sharp spectral features such as emission and absorption lines.
    Despite advancements in multi-object spectrographs, spectroscopic measurements are expensive and time-consuming and we can only collect spectra for a small fraction of the galaxies that can be imaged by  modern surveys, such as the Dark Energy Survey (DES; \citealt{TheDarkEnergySurveyCollaboration2005}) and the Kilo-Degree Survey (KiDS; \citealt{DeJong2013a}).
    This problem will only increase in magnitude as the next generation of surveys, such as the Vera Rubin Observatory Legacy Survey of Space and Time (LSST; \citealt{LSSTScienceCollaboration2009}) and the Wide-Field Infrared Survey Telescope (WFIRST; \citealt{Green2012}), image orders of magnitude more galaxies at fainter magnitudes than are present in current data sets.
    As a result, rather than rely on spectroscopic redshifts (spec-z's), modern surveys increasingly rely on photometric redshifts (photo-z's; see \citealt{Salvato2019} for a review).

    Photo-z's are estimates of galaxy redshifts derived from changes in the colors of galaxies as their spectral energy distributions (SED's) redshift through a series of broadband filters.
    This estimation is typically done using one of two approaches: machine learning (ML) or template fitting (see e.g. \citealt{Schmidt2020} for an evaluation of many examples of the two).

    Machine learning approaches train on a data set of photometry with spec-z's and attempt to directly learn an empirical relationship between galaxy colors and redshift (e.g. \citealt{Connolly1995}, \texttt{TPZ} \citealt{Kind2013}, \texttt{FlexZBoost} \citealt{Izbicki2017}, \texttt{CMNN} \citealt{Graham2018a}).
    Once trained, they can predict galaxy redshifts given photometry alone.
    The advantage of ML methods is that the effects of dust, galaxy evolution, and other relevant variables are encoded in the training set and thus it is possible for ML methods to account for these in the derived mapping from colors to redshift if the data encapsulate these effects.
    The success of this mapping depends on the  choice and complexity of the ML model and the corresponding hyperparameters.
    The downside of ML methods is that their success relies on how representative and well-controlled the training set is, and that they are unable to extrapolate beyond that set.

    Template fitting photo-z estimators (e.g. \texttt{LePhare} \citealt{Arnouts1999}, \bpz\ \citealt{Benitez2000a}, \texttt{EAZY} \citealt{Brammer2008}) work on the assumption that galaxy photometry are sampled from a relatively small set of underlying spectral types, characterized by the eponymous SED templates. 
    These estimators calculate photo-z's by selecting the template and redshift with simulated fluxes most similar to the observed fluxes. 
    In order for this method to work, the underlying SED templates from which the galaxies are sampled must be known. 
    Common methods for generating these templates include simulating galaxy SED's from spectral synthesis models, e.g. \citet{BruzualA.1993a}, and deriving templates from the observed spectra of local galaxies, e.g. \citet{Benitez2004a}. 

    The primary advantage of template fitting methods is that it is not limited to the bounds of a training set. 
    A key limitation is that they do not guarantee that the SED templates will span the full distribution of galaxy spectra in a given data set, nor that they will properly account for the effects of dust, or spectral evolution.
    In addition, spectral synthesis models are only able to generate spectra with a discrete set of physical parameters (e.g. temperature, age, metallicity), and obtaining real galaxy spectra is expensive, especially at the redshifts and magnitudes that will be observed by LSST.

    Several previous works have attempted to combine the advantages of these two approaches by deriving SED templates from a photometric training set, and then using the derived templates for photo-z estimation (\citealt{Budavari2000b,Csabai2000}). These approaches
    leverage a large set of galaxy photometry, which amount to low resolution spectra, to sample a smaller set of SED templates across a broad range of rest wavelengths.
    This effectively over-samples the template SED's, allowing us to reconstruct spectral features at a resolution much higher than that of the broadband filters used to measure the photometry.
    This  is analogous to the Drizzle technique used to reconstruct higher resolution images for the Hubble Space Telescope (HST; \citealt{Fruchter2002}) and the reconstruction of SED's using differential chromatic refraction (DCR; \citealt{Lee2019}).

    This template learning approach retains the physical motivation and extensibility of the template fitting method, while taking advantage of learning the systematics and confounding variables implicit in the training set.
    In addition, it opens the possibility of learning a smooth continuum of galaxy spectra, in contrast to the discrete set offered by the limited galaxy observations and galaxy modeling codes. 

    While previous works attempt to learn galaxy templates from data using a set of eigenspectra, we adapt the algorithm of \citet{Budavari2000b} to directly learn a set of templates from the data.
    We extend these earlier works by applying our methods to a large data set of 102,476 galaxies with spec-z's and photometry in 19 bands.
    In this manner we are able to learn a variable number of SED templates with clear spectral features, and with simple postprocessing, we are able to further reconstruct emission lines in the bluest templates.

    We show that templates can be learned from scratch or as perturbations of pre-existing templates.
    We use these learned templates to estimate photo-z's with \bpz\ and find that the training reduces the bias and scatter of the redshift estimates, with little impact on the fraction of catastrophic outliers.
    In addition, we find that both bias and scatter decrease with the number of SED templates used in the photo-z estimation.

    The outline of the paper is as follows:
    in Section \ref{sect:template_training} we describe the template training algorithm, including how to match photometry to templates, how to perturb templates to better match the photometry, and how to select the hyperparameters for training.
    In Section \ref{sect:data}, we describe the spec-z and photometric data sets used in the template training and redshift estimation.
    In Section \ref{sect:application}, we apply the template training algorithm to sets of naive templates and to a pre-existing set of templates derived from galaxy observations and spectral synthesis models.
    We discuss the performance of the algorithm including its convergence, and the accuracy of the reconstructions.
    In Section \ref{sect:photoz}, we use our templates to estimate photo-z's for a training set of galaxies and analyze the results.
    We discuss our results and future goals in Section \ref{sect:discussion} and conclude in Section \ref{sect:conclusion}.
    
\section{Template Training Algorithm}

    \label{sect:template_training}

    In this section, we will present an approach for learning SED templates directly from broadband photometry, using a modified version of the algorithm developed in \citet{Budavari2000b}. 
    If we assume that the galaxies in our data set are sampled from a small set of underlying spectra, the SED templates, and we know the spectroscopic redshift for each galaxy, we can shift the photometry to the restframe and treat each observation of a redshifted galaxy as a \textit{restframe observation} of one of the templates with a different set of effective filters. 
    With a large enough data set, the wavelengths of the effective filters will overlap substantially. 
    This over-sampling allows us to recover higher resolution features in the templates, even though the data are low resolution observations of different galaxies.

    Let us assume we have a set of SED templates as a starting point, which can represent rudimentary guesses and need not resemble true galaxy spectra. 
    In the first part of this section, we describe a method by which we create a training set of broadband photometry for each template from a large data set of galaxy photometry. 
    In the second part, we derive the perturbation algorithm that is used to train each SED template on its corresponding photometry set. 
    The full training algorithm is an expectation maximization that consists of iterating these two steps: matching photometry to templates, and perturbing templates to better match the photometry.
    This process is iterated until the SED templates converge. 
    In the final part, we discuss a heuristic for selecting the training hyperparameters.

    \subsection{Matching Photometry Sets}
    \label{sect:training_sets}
            
    Assume we have a set of naive SED templates and a large set of observed fluxes, $\{f_m\}$, with known spectroscopic redshifts, $z_m$. 
    Our goal is to train each template on an appropriate subset of the $\{f_m\}$, so that the naive templates better represent the colors of the galaxies. 
    To assemble these training sets, we consider subsets $\{f_n\} \subset \{f_m\}$, corresponding to the observed fluxes of a single galaxy at redshift $z$, where the subscript $n$ denotes different filters. 
    We compare these observed fluxes with the template fluxes $\{\hat{f}_n\}$, where
    \begin{align}
        \hat{f}_n &= \int S\left(\frac{\lambda}{1+z}\right) R^n(\lambda) d\lambda, \label{eq:calc_flux1}
    \end{align}
    $S(\lambda)$ is an SED template, and $R^n(\lambda)$ is the normalized response function of the filter used to measure the flux $f_n$.
    For photon counting detectors,
    \begin{align}
        R(\lambda) = \frac{\lambda \, T(\lambda)}{\int \lambda \, T(\lambda) d\lambda},
    \end{align}
    where $T(\lambda)$ is the system response function that captures the transmittance of the atmosphere and the response of the detector \citep{Bessell2005}.

    The observed fluxes are assigned to the template whose colors are most similar, which is determined by normalizing the observed and template fluxes in the same band and picking the template that minimizes the squared differences of the fluxes. 
    The normalization band is chosen by selecting the band for which the ratio $\hat{f}_n / f_n$ is the median of the flux ratios for that galaxy. 
    By performing this matching and renormalization for each galaxy in the photometry set, we associate a subset of the galaxies (and the corresponding photometry) to each template.

    Examining how the galaxies are assigned to the individual templates is helpful in selecting the initial set of templates.
    The initial templates should be chosen so that the matching algorithm roughly divides the galaxies by their colors.
    It is also important that each set contains a sufficient number of fluxes distributed across the wavelengths of interest, as the perturbation algorithm derived in the next section relies on over-sampling to reconstruct higher resolution features of the SED templates.

    \subsection{The Perturbation Algorithm}
    \label{sect:perturbation}

    Assume we have a set of photometry, $\{f_n\}$, which constitute observations of the same underlying SED template, $S(\lambda)$, at various known redshifts, $z_n$. 
    These observed fluxes should approximately match the template fluxes calculated via Equation \ref{eq:calc_flux1}. 
    However, we can also calculate the template fluxes by imagining that we are observing the template in its rest frame using a set of effective, blueshifted filters:
    \begin{align}
        \hat{f}_n &= \int S(\lambda) \, R^n[(1+z_n)\lambda] \, d[(1+z_n)\lambda] \\
                  &= \sum_k s_k \, r_{k'}^n \Delta\lambda_{k'}, \label{eq:calc_flux2}
    \end{align}
    where in the second line $s_k$ and $r_k^n$ are the discrete representations of $S(\lambda)$ and $R^n(\lambda)$ respectively, parameterized by the wavelength bins $\{\lambda_k\}$ with widths $\{\Delta\lambda_k\}$.
    Primed indices indicate redshifted wavelengths, i.e. $\lambda_{k'} = (1+z_n) \lambda_k$ and $\Delta\lambda_{k'} = (1+z_n)\Delta\lambda_k$.

    We wish to perturb the template so that the template fluxes, $\hat{f}_n$, better match the observed fluxes, $f_n$. 
    Letting $\hat{s}_k$ be a new template resulting from a perturbation of $s_k$, we define the cost function (\citealt{Budavari2000b} Equation 7):
    \begin{align}
        \chi^2 =
        \sum_n \frac{1}{\sigma_n^2}(\hat{f}_n(\{\hat{s}_k\}) - f_n)^2 + 
        \sum_k \frac{1}{\Delta_k^2}(\hat{s}_k - s_k)^2. \label{eq:cost_function}
    \end{align}
    The optimum perturbation is found via a multidimensional minimization of the cost function. 
    The first term in Equation \ref{eq:cost_function} penalizes differences between the observed fluxes and the perturbed template fluxes, weighted according to $\sigma_n$ (the fractional error of the measured flux).
    The second term in Equation \ref{eq:cost_function} penalizes large perturbations, weighted by the hyperparameters $\Delta_k$. 
    This parameter controls learning rate and also helps stabilize the results. 
    See the next section for more details. 

    We follow \citet{Budavari2000b} by introducing the simplifying perturbation and constant terms: 
    \begin{align}
        \begin{gathered}
            \xi_k = \hat{s}_k - s_k \\
            g_n = f_n - \sum_k s_k \, r_{k'}^n \Delta\lambda_{k'}.
        \end{gathered}
    \end{align} 
    Then, we have:
    \begin{align}
        \chi^2 = 
        \sum_n \frac{1}{\sigma_n^2} \left( g_n - \sum_k \xi_k \,  r_{k'}^n \Delta\lambda_{k'} \right)^2 +
        \sum_k \frac{\xi_k^2}{\Delta_k^2},
    \end{align}
    which can be analytically minimized:
    \begin{align}
        \frac{\partial \chi^2}{\partial \xi_l} = 0 \implies \sum_k M_{lk} \tilde{\xi}_k = \nu_l.
    \end{align}
    The matrix $M$ and vector $\nu$ are defined
    \begin{align}
        \begin{gathered}
            M_{lk} = \sum_n \frac{1}{\sigma_n^2} (r_{l'}^n \Delta\lambda_{l'}) (r_{k'}^n \Delta\lambda_{k'}) + \frac{\delta_{lk}}{\Delta_k^2}, \\
            \nu_l = \sum_n \frac{g_n}{\sigma_n^2} (r_{l'}^n \Delta\lambda_{l'}),
        \end{gathered}
    \end{align}
    where $\delta_{lk}$ is the Kronecker delta.
    One can then solve for $\tilde{\xi}$. 
    The perturbed spectrum is then $\hat{s}_k = s_k + \tilde{\xi}_k$. 

    \begin{figure}
        \centering
        \includegraphics{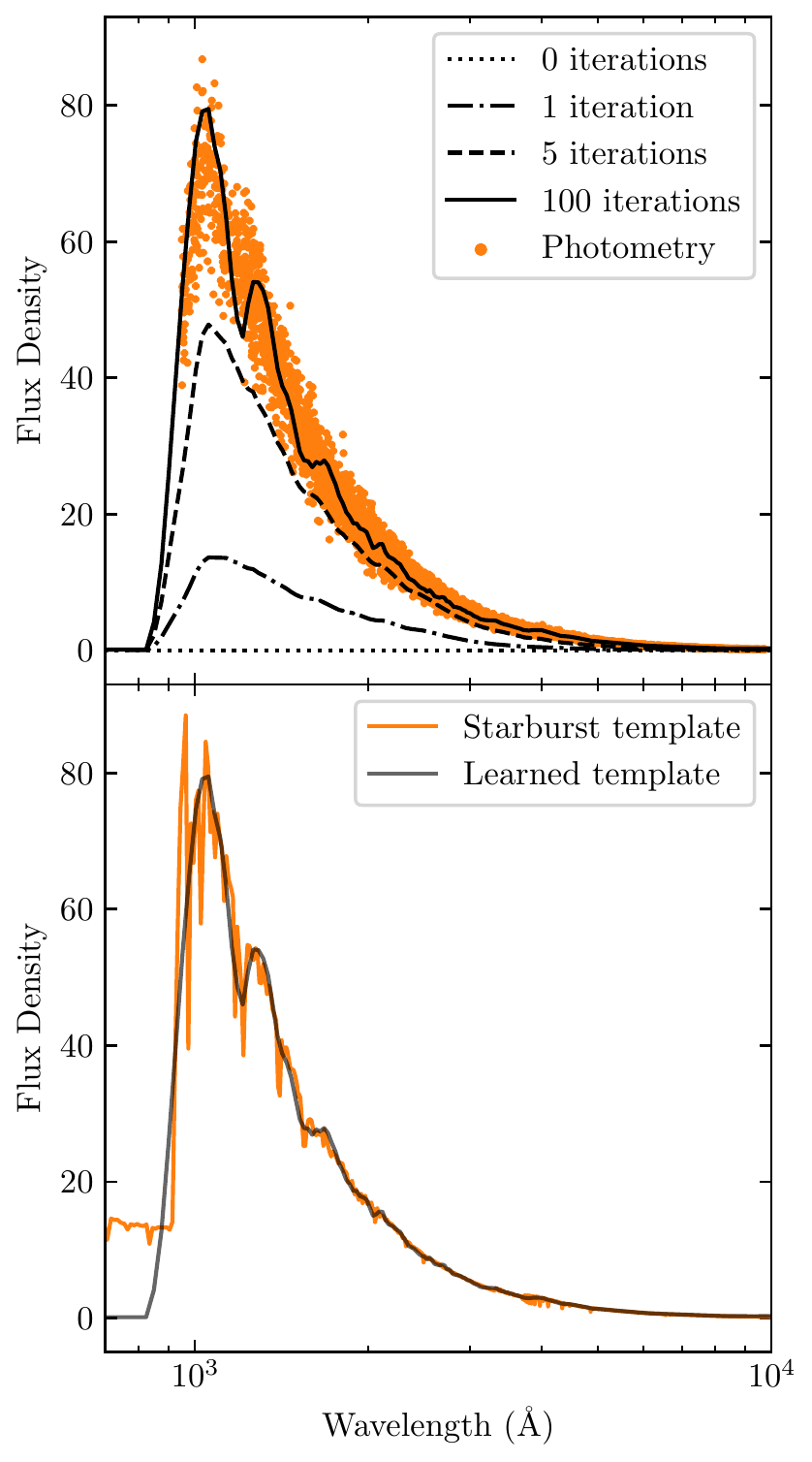}
        \caption{Perturbing a naive template, in this case a flat line, to better match a photometry set. Top: the orange points are simulated observations of the 5Myr starburst template from \citet{Coe2006a} at 1,000 random redshifts in the range z=0 to z=3 using the $ugrizY$ filters listed in Table \ref{tab:filters}. The simulated photometry has 10\% Gaussian error. The template is shown after various stages of the training. Bottom: the learned template is plotted with the original starburst template.}
        \label{fig:pert_ex}
    \end{figure}

    Iterating the perturbation changes the shape of the template SED to better match the measured photometry, as shown in \citet{Budavari2000b}. 
    An example of this process can be seen in Figure \ref{fig:pert_ex}. 
    Fluxes in the $ugrizY$ filters listed in Table \ref{tab:filters} were calculated for a starburst galaxy template at 1000 random redshifts $z < 3$. 
    Starting with an $S(\lambda) = 0$ template SED, the perturbation algorithm is applied iteratively. 
    After 100 iterations, the trained template closely matches the original template in the wavelength range for which photometry exists. 
    While the trained template is a smoothed version of the original, high resolution features have been recovered, despite the relatively low resolution of the filters. 
    In practice, higher $\Delta_k$ can be chosen so that fewer iterations are required in the training; a lower value was chosen here so that the effects of successive iterations can be more clearly seen.
    See Section \ref{sect:hyperparameters} for further discussion of selecting the hyperparameters.

    The perturbation algorithm changes the shape of the template SED's so that re-running the photometry matching will now result in different subsets of galaxies assigned to each template.
    The full training algorithm is iterated until the SED templates converge.

    \subsection{Selecting Hyperparameters}
    \label{sect:hyperparameters}

    The success of the training algorithm depends on the chosen hyperparameters.
    The first is the number of templates. 
    As discussed in Section \ref{sect:training_sets}, this choice can be made by using the photometry matching algorithm and choosing the appropriate number of templates to approximately separate out the different spectral shapes displayed in the photometry.
    For further discussion of how the number of templates affects photo-z results, see Section \ref{sect:photoz_results}.

    The rest of the hyperparameters consist of the set of $\Delta_k$.
    These parameters, which set the relative weighting of the regularization term in Equation \ref{eq:cost_function}, determine the stability and speed of the training algorithm.
    If the $\Delta_k$ are too large, training will be very slow and a large number of iterations will be required. 
    If the $\Delta_k$ are too low, the training becomes unstable and the final templates will be over-fit.
    Here we present a heuristic for selecting an appropriate value to balance these two extremes.

    For the work presented below, the index $k$ is dropped, so that $\Delta \equiv \Delta_k$ has a single value for each training set that is independent of wavelength. 
    In choosing the appropriate value of $\Delta$ for each training set, it is desirable to select a value that corresponds to a constant ratio, $w$, of the flux and regularization terms in Equation \ref{eq:cost_function}. 
    The necessary value of $\Delta$ will vary by training set, as the number of terms in the sum over fluxes (i.e. the sum over $n$ in Equation \ref{eq:cost_function}) will vary by training set. 
    To this end, we make the following approximation:
    \begin{align}
        \frac{\sum_k \left(\hat{s}_k - s_k \right)^2}{\sum_n \left(\hat{f}_n - f_n \right)^2} \sim \frac{N_k}{N_n},
    \end{align}
    where $N_k \equiv \sum_k$ and $N_n \equiv \sum_n$. 
    This permits the following approximation of the ratio $w$: 
    \begin{align}
        w = \frac{\sum_k \Delta^{-2} \left(\hat{s}_k - s_k \right)^2}{\sum_n \sigma_n^{-2} \left(\hat{f}_n - f_n \right)^2} \sim \frac{N_k/\Delta^2}{N_n/\Bar{\sigma}^2},
    \end{align}
    where $\Bar{\sigma} = \sum_n \sigma_n/N_n$. 
    Then, for a desired ratio $w$, the requisite $\Delta$ can be approximated:
    \begin{align}
        \Delta \simeq \Bar{\sigma} \sqrt{\frac{N_k}{w N_n}}.
    \end{align}
    In practice, we have found that $w = \mathcal{O}(1)$ works well.
    The results of the training are relatively robust to the selection of $w$, in that changing $w$ by, for example, a factor of 2 yields similar results.
        
\section{Data}

    \label{sect:data}

    We collect a set of galaxy spectroscopic redshifts, paired with broadband photometry, from various surveys to test our training algorithm.
    Our set consists of 102,476 galaxies with redshifts $z < 4.54$ and $i$-band magnitudes\footnote{The $i$-band magnitudes quoted in this section denote the magnitude in one of $i$, $i_2$, $I$, or $i^+$ as listed in Table \ref{tab:filters}. For galaxies with photometry in multiple $i$-bands, the magnitude used is the first to appear in that list.} in the range $13.8 < i < 25.7$.
    For all surveys, we use galaxies with highly reliable spec-z's, photometry in one of the $i$-bands, and photometry in at least three bands with signal-to-noise ratio SNR $> 20$.
    The entire data set is summarized in Table \ref{tab:data_sets}, the filters used to measure the photometry are listed in Table \ref{tab:filters}, and the redshift distributions are shown in Figure \ref{fig:redshift_dist}.

    \begin{deluxetable*}{l R C C C C C C l}
        \centerwidetable
        \tablecaption{Summary of the Spectrophotometric Data Sets \label{tab:data_sets}}
        \tablehead{\colhead{Data Set} & \colhead{$N_\text{gal}$} & \colhead{$f_\text{gal}$} & \colhead{$z_\text{mean}$} & \colhead{$z_\text{max}$} & \colhead{$i$-band range} & \colhead{$i_\text{mean}$} & \colhead{$\Bar{\sigma}_i$} & \colhead{Link to Catalog}}
        \startdata
            zCOSMOS  &  14298 & 0.14 & 0.57 & 2.52 & 16.87 \leq i \leq 24.18 & 21.19 & 0.022 & \url{http://cesam.lam.fr/hstcosmos/} \\
            VVDS     &   6915 & 0.07 & 0.67 & 4.54 & 13.84 \leq i \leq 24.97 & 20.86 & 0.014 & \url{https://cesam.lam.fr/vvds/index.php} \\
            VIPERS   &  69415 & 0.68 & 0.70 & 2.15 & 17.66 \leq i \leq 23.08 & 21.38 & 0.017 & \url{http://vipers.inaf.it:8080/} \\
            DEEP2/3  &  10695 & 0.10 & 0.71 & 1.91 & 15.30 \leq i \leq 25.36 & 21.42 & 0.020 & \url{http://d-scholarship.pitt.edu/36064/} \\
            3D-HST   &   1153 & 0.01 & 1.46 & 3.32 & 19.10 \leq i \leq 25.74 & 23.56 & 0.027 & \url{http://d-scholarship.pitt.edu/36064/} \\
            \midrule
            Training &  81980 & 0.80 & 0.69 & 4.54 & 13.84 \leq i \leq 25.74 & 21.32 & 0.018 & \\
            Test     &  20496 & 0.20 & 0.69 & 3.61 & 16.46 \leq i \leq 25.69 & 21.34 & 0.018 & \\
            \midrule
            Total    & 102476 & 1.00 & 0.69 & 4.54 & 13.84 \leq i \leq 25.74 & 21.33 & 0.018 & \\
        \enddata
        \tablecomments{$N_\text{gal}$ is the total number of galaxies in the set, $f_\text{gal}$ is the fraction of all galaxies in the set, and $\Bar{\sigma}_i$ is the mean fractional flux error in the $i$-band.}
    \end{deluxetable*}

    \begin{deluxetable}{L c c R R}
        \tablecaption{List of the Broadband Filters \label{tab:filters}}
        \tablehead{\colhead{Filter} & \colhead{Telescope} & \colhead{Instrument} & \colhead{$\lambda_0$} & \colhead{$W_{\text{eff}}$}}
        \startdata
            NUV & GALEX  &         &  2343.1 &  767.3 \\
            u   & CFHT   & Megacam &  3817.7 &  525.4 \\
            B   & CFHT   & CFH12k  &  4342.5 &  873.6 \\
            B_J & Subaru & Suprime &  4478.4 &  763.9 \\
            g^+ & Subaru & Suprime &  4808.5 & 1043.1 \\
            g   & CHFT   & Megacam &  4899.9 & 1293.8 \\
            V   & CFHT   & CFH12k  &  5393.7 &  882.7 \\
            V_J & Subaru & Suprime &  5493.0 &  862.4 \\
            r   & CHFT   & Megacam &  6278.2 & 1120.2 \\
            r^+ & Subaru & Suprime &  6314.8 & 1211.4 \\
            R   & CFHT   & CFH12k  &  6603.5 & 1138.5 \\
            i_2 & CHFT   & Megacam &  7584.5 & 1409.4 \\
            i   & CHFT   & Megacam &  7676.6 & 1307.6 \\
            i^+ & Subaru & Suprime &  7709.1 & 1361.7 \\
            I   & CFHT   & CFH12k  &  8277.3 & 1816.7 \\
            z   & CHFT   & Megacam &  8857.6 & 1040.1 \\
            z^+ & Subaru & Suprime &  9054.5 & 1012.3 \\
            Y   & Subaru & Suprime & 10216.0 &  996.2 \\
            J   & UKIRT  & WFCAM   & 12508.5 & 1476.8 \\
        \enddata
        \tablecomments{ Mean wavelength, $\lambda_0 = \int \lambda R(\lambda) d\lambda$, and effective width, $W_\text{eff} = \text{Max}[R(\lambda)]^{-1}$, are given in angstroms. Filters are listed in order of increasing $\lambda_0$. The response functions for each filter were obtained from the Spanish Virtual Observatory (SVO) Filter Profile Service.}
    \end{deluxetable}

    \medskip
    \subsection{zCOSMOS-\textit{bright}}

    zCOSMOS \citep{Lilly2009a} is a redshift survey of 1.7 $\text{deg}^2$ of the COSMOS field, conducted with the VIMOS spectrograph mounted on the European Southern Observatory's (ESO) Very Large Telescope (VLT).
    The survey is divided into two parts, \textit{bright} and \textit{deep}. 
    We make use of the former, consisting of approximately 20,000 galaxies with redshifts $z < 1.2$.
    We use galaxies recommended in the ESO data release description\footnote{\url{https://www.eso.org/sci/observing/phase3/data_releases/zcosmos_dr3_b2.pdf}}, determined to have $99\%$ spectroscopic verification (i.e. \texttt{zflag} = 3.x, 4.x, 2.5, 2.4, 1.5, 9.5, 9.3, 18.5, 18.3).

    The zCOSMOS redshifts are matched to photometry from \citet{Ilbert2009}.
    The photometry is measured from the ultraviolent to the near-infrared in 11 broadband filters: $NUV$ on GALEX \citep{Martin2005a}, $u$ and $i$ on CFHT-Megacam, $B$ and $V$ on CFHT-CFH12k, $g^+$, $r^+$, $i^+$, and $z^+$ on Subaru, and $J$ on UKIRT.
    The final set consists of 14,298 galaxies with redshifts $z < 2.52$ and $i$-band magnitudes in the range $16.9 < i < 24.2$.

    \subsection{VVDS}

    The VIMOS VLT Deep Survey (VVDS, \citealt{LeFevre2013b}) is a redshift survey consisting of three component surveys: \textit{Wide}, \textit{Deep}, and \textit{Ultra-Deep}. 
    The Wide survey covers 8.7 $\text{deg}^2$, with approximately 25,000 galaxies in the range $17.5 < i < 22.5$; the Deep survey covers 0.74 $\text{deg}^2$, with approximately 11,000 galaxies in the range $17.5 < i < 24$; the Ultra-Deep survey covers 512 $\text{arcmin}^2$, with approximately 900 galaxies in the range $23 < i < 24.75$.
    We use redshifts with quality flags 3 and 4, indicating a 98\% spec-z confidence.
    The photometry was measured in nine filters: $u,g,r,i,z$ on CFHT-Megacam \citep{Hudelot2012} and $B,V,R,I$ on CFHT-CFH12k \citep{LeFevre2004}.
    The final set contains 6,915 galaxies out to redshifts $z < 4.5$, with magnitudes $ 13.8 < i < 25.0$.

    \subsection{VIPERS}

    The VIMOS Public Extragalactic Redshift Survey (VIPERS, \citealt{Scodeggio2018a}) is a dense, large-volume redshift survey focusing on redshifts $0.5 < z < 1.2$.
    We use VIPERS galaxies with spec-z's reliable at the 95\% confidence level (\texttt{zflag} = 2.X, 3.X, 4.X), and with \texttt{photoMask} and \texttt{spectroMask} = 1.
    The redshifts are matched to photometry measured in $NUV$ on GALEX \citep{Martin2005a}, and $u,g,r,i_2,i,z$ on CFHT-Megacam\textsuperscript{\ref{ft:i2}} \citep{Hudelot2012}. 
    The final set contains 71,951 galaxies with redshifts $z < 2.15$ and magnitudes $17.7 < i < 23.3$. 

    \subsection{DEEP2 and DEEP3}

    DEEP2 and DEEP3 are redshift surveys conducted with the DEIMOS spectrograph on the Keck 2 telescope.
    DEEP2 \citep{Newman2013b} consists of four fields; we use galaxies from the first field in the Extended Groth Strip (EGS), which had no redshift pre-selection.
    DEEP3 \citep{Cooper2011} expanded on the DEEP2 survey of the EGS.
    Redshifts from these surveys are matched with aperture-corrected photometry provided by \citet{Zhou2019a}.
    We use galaxies with CFHTLS flag 0, SExtractor flags less than 4 in every band, and redshift quality flag $\geq 3$.
    Photometry was measured in $u,g,r,i_2,i,z$ on CFHT-Megacam\footnote{The $i_2$ band is the replacement to the Megacam $i$-band installed in 2007. This filter is named $y$ in the CFHTLS catalogues \citep{Hudelot2012}, but we follow \citet{Zhou2019a} in naming it $i_2$ to avoid confusion with the longer $y$ bands used in Subaru and LSST. \label{ft:i2}} and $Y$ on Subaru \citep{Miyazaki2002}.
    The final set contains 10,695 galaxies with redshifts $z < 1.91$ and magnitudes $15.3 < i < 25.74$.

    \subsection{3D-HST}

    In addition to the spectroscopic surveys above, we include grism redshifts from the 3D-HST survey \citep{Newman2013b,Momcheva2016b}.
    Redshifts for this survey were analyzed and matched with aperature-corrected photometry by \citet{Zhou2019a}.
    We select the galaxies with CFHTLS flag 0, SExtractor flags less than 4 in every band, and the flag \texttt{use\_zgrism1} = 1.
    For galaxies in both the DEEP2/3 and 3D-HST sets, we use DEEP2/3 redshifts instead.
    Photometry was measured in $u,g,r,i_2,i,z$ on CFHT-Megacam and $Y$ on Subaru.
    After these cuts, the 3D-HST set contains 1,153 galaxies with redshifts $z < 3.32$ and magnitudes $23.6 < i < 25.7$.

    \begin{figure}
        \centering
        \includegraphics{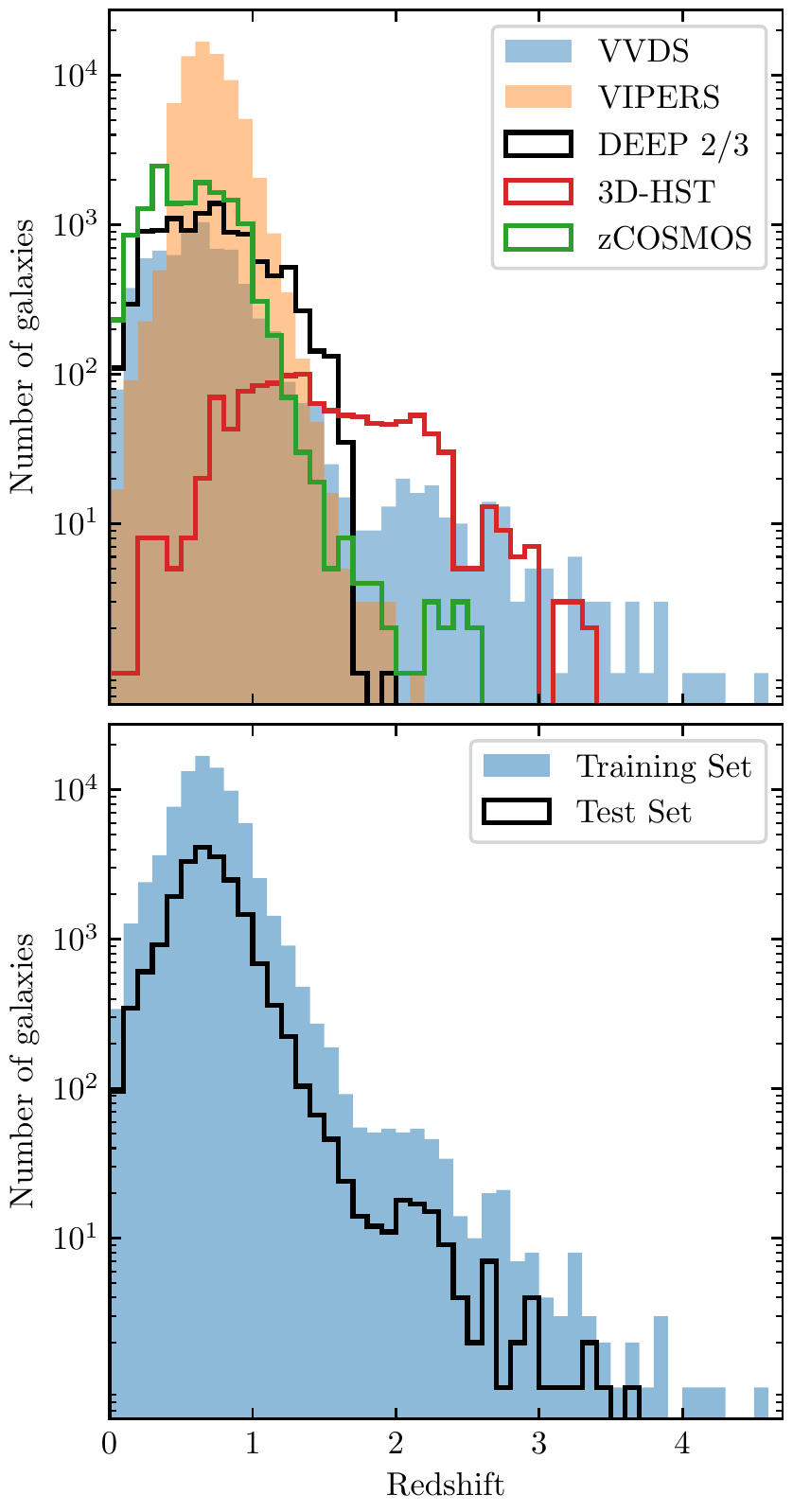}
        \caption{Redshift distribution of the galaxy surveys. The top panel shows the distributions of each of the constituent surveys. The bottom panel shows the redshift distributions of the training and test sets used for template training and photo-z estimation respectively.}
        \label{fig:redshift_dist}
    \end{figure}

\section{Application to Data}

    \label{sect:application}

    Using the training algorithm described in Section \ref{sect:template_training}, we will learn galaxy SED templates directly from the broadband photometry described in Section \ref{sect:data}.
    We divide the data set into a training and test set, consisting of random 80\% and 20\% samples respectively of the entire data set.
    The training set will be used to train the SED templates, while the test set will be used to evaluate the learned templates via photo-z estimation (see Section \ref{sect:photoz}).
    The training set consists of 81,980 galaxies, with mean redshift $z_\text{mean} = 0.69$, max redshift $z_\text{max} = 4.54$, and magnitudes $13.8 < i < 25.7$.
    A full summary of the set can be seen in Table \ref{tab:data_sets}, and the redshift distribution can be seen in Figure \ref{fig:redshift_dist}.

\begin{figure*}
        \centering
        \includegraphics{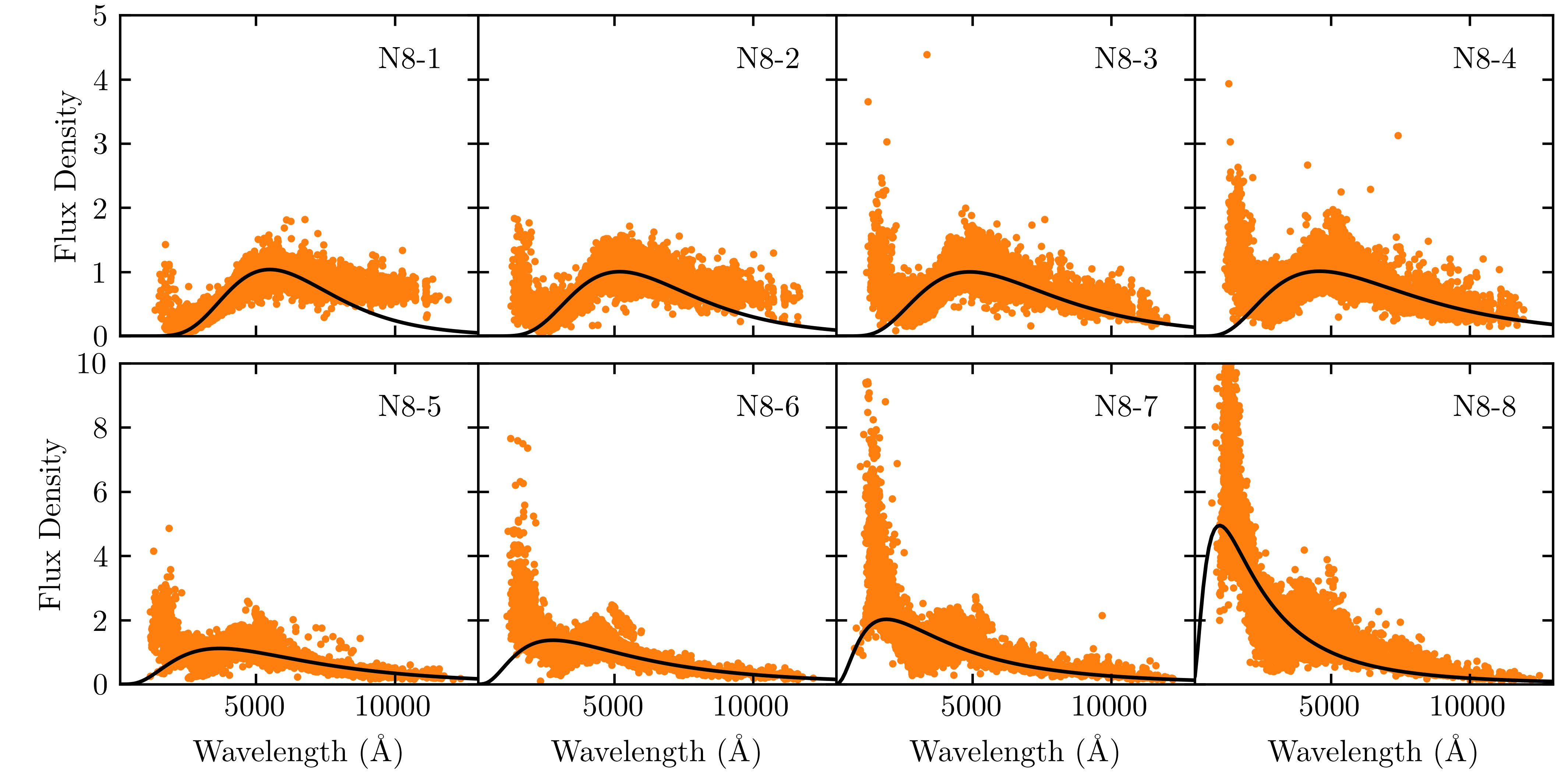}
        \caption{The untrained N8 templates (black lines) with their corresponding photometry sets (orange points), generated with the algorithm described in Section \ref{sect:training_sets}. N8-1 is the reddest template, with each successive template getting bluer.}
        \label{fig:N8_untrained}
    \end{figure*}

    Eight naive templates were chosen to represent the underlying SED shapes of the photometry set according to the principles described at the end of Section \ref{sect:training_sets}.
    We chose the number eight to allow a direct comparison to the standard template set described below.
    They are ``naive'' because they are simply chosen by eye to roughly divide the photometry into groups by spectral shape, but otherwise are not based on any theoretical models or observed SED's.
    Each of the naive templates is a log-normal function,
    \begin{align}
        S(\lambda) \propto \frac{1}{\lambda} \exp{\left[ -\frac{1}{2\eta^2} \left( \ln{\frac{\lambda}{\text{mode}(\lambda)}}-\eta^2 \right)^2 \right]},
    \end{align}
    normalized at $\lambda = 5000$ \AA, with $\text{mode}(\lambda)$ in the range $1000$ to $5500$ \AA\  and $\eta$ in the range $0.35$ to $0.9$. 
    The templates extend to $15000$ \AA\ with 100 \AA\ resolution.
    These eight templates (hereafter N8) can be seen together with with their original training sets in Figure \ref{fig:N8_untrained}.

    The training algorithm with $w=0.5$ is applied to the N8 templates.
    The convergence of the templates is evaluated via the weighted mean square error,
    \begin{align}
        \text{wMSE} = \sum_n \frac{1}{\sigma_n^2}(\hat{f}_n(\{\hat{s}_k\}) - f_n)^2.
    \end{align}
    Each template is perturbed until the change in wMSE is less than 3\%, which was chosen empirically to balance sufficient template reconstruction and the algorithm's runtime.
    When every template has converged to its current photometry set, new photometry sets are generated.
    Only those templates whose new photometry sets result in a greater than 3\% change in wMSE resume perturbation with their new sets.
    This process is iterated until no template has a new photometry set that results in a greater than 3\% change in wMSE.
    This indicates that the photometry is sorted into distinct sets, and that further perturbation is unlikely to improve the photometry-matching results.

\begin{figure*}
        \centering
        \includegraphics{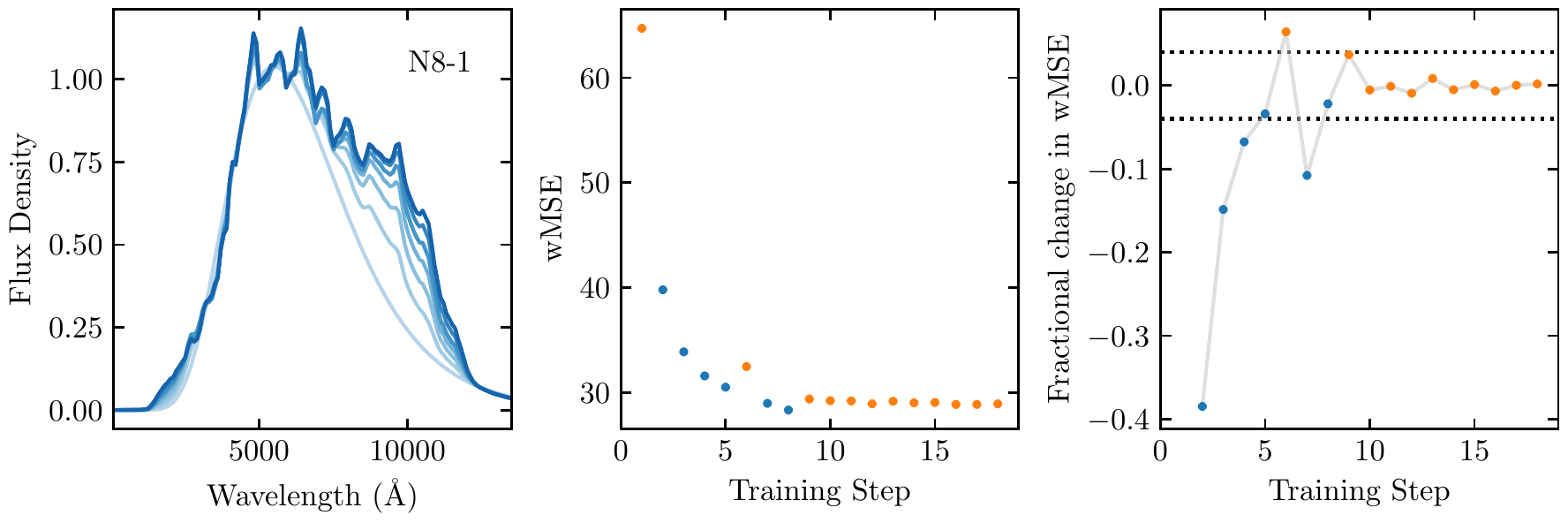}
        \caption{Training of N8-1. Left: the initial (light blue) N8-1 template is iteratively perturbed to better represent the colors of its photometry set. The final (dark blue) template is redder and has more structure. Middle: wMSE of the N8-1 template throughout the training process. Orange points represent the wMSE after a photometry matching stage, while blue points represent the wMSE after a perturbation. Right: fractional change in the wMSE. Orange points represent the fractional change due to a new photometry matching stage, while blue points represent a fractional change due to a perturbation. The dotted black lines show the $\pm 0.03$ cutoff. When a perturbation results in a fractional change of magnitude less than 0.03, perturbation is halted and new photometry is matched. After the sixth photometry match, the template is not perturbed because it already sufficiently matches the photometry.}
        \label{fig:training}
    \end{figure*}

    The progress of the training algorithm is shown in Figure \ref{fig:training} for the template N8-1.
    The left panel shows the progress of the perturbation algorithm as it deforms the originally smooth N8-1 template to better match the colors of the matched photometry sets.
    In particular, N8-1 becomes redder and acquires higher resolution structure, which will be discussed below.
    The middle panel shows the wMSE and the right panel shows the fractional change in the wMSE throughout the training.
    Orange points indicate values after a photometry-matching stage, and blue points indicate values after a perturbation.
    You can see that the wMSE drops as the template is perturbed, and perturbation continues until the magnitude of the fractional change in wMSE drops below 0.03, indicated by the dotted black lines in the right panel.
    Once this occurs, new photometry is matched, resulting in an increase in wMSE.
    This process is iterated, with fewer and fewer perturbations needed per iteration.
    Eventually, all of the points are orange, indicating that after each new photometry matching, N8-1 is not perturbed, as it already sufficiently matches its photometry set.

\begin{figure*}
        \centering
        \includegraphics{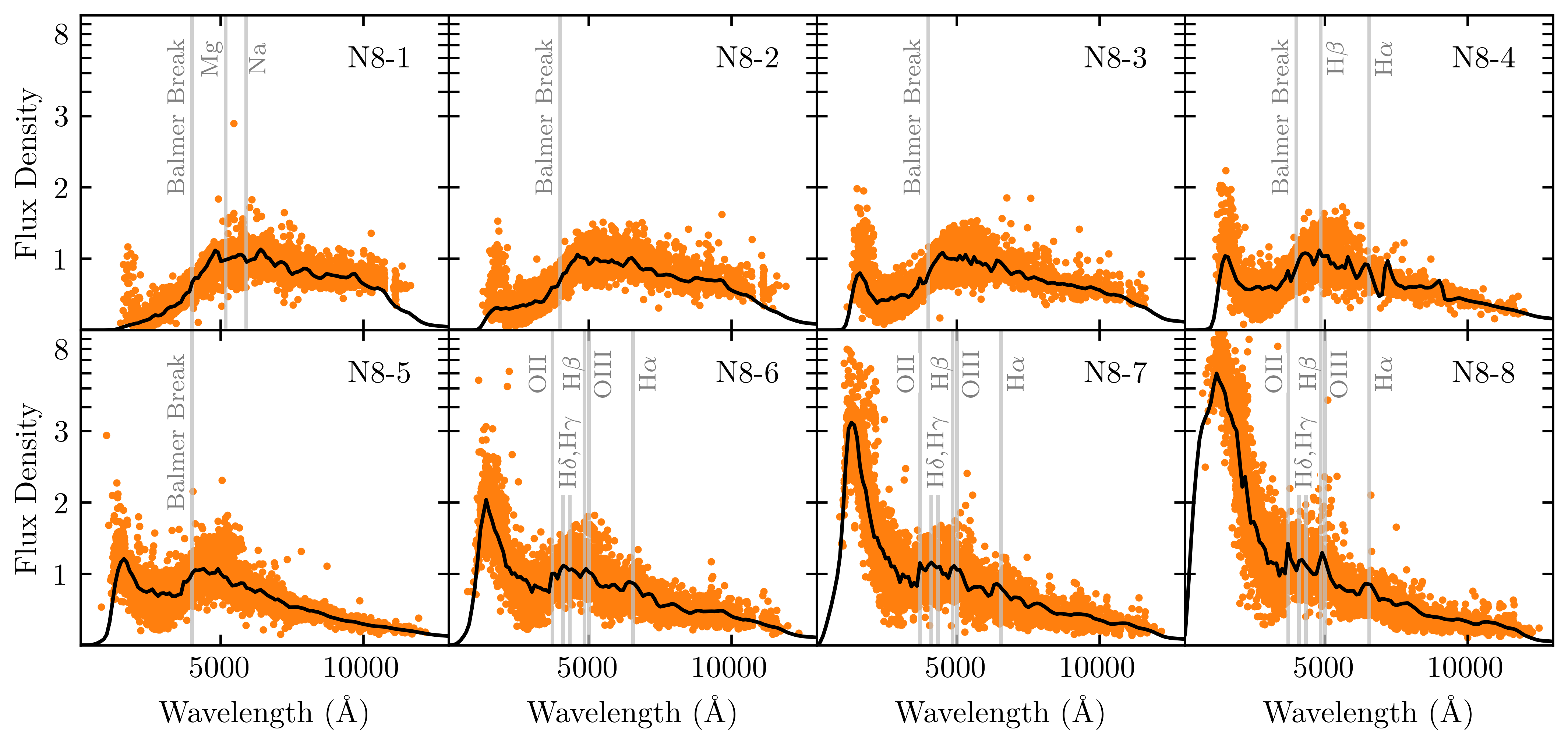}
        \caption{The trained N8 templates (black lines) with their final photometry sets (orange points). N8-1 is the reddest template, with each successive template getting bluer. The templates now more closely resemble physical galaxy spectra, and have acquired structure at a higher resolution than the broadband templates. The Balmer break, Mg and Na absorption lines, and H$\alpha$, H$\beta$, H$\gamma$, H$\delta$, OII, and OIII emission lines are labeled in gray.}
        \label{fig:N8_trained}
    \end{figure*}

    The training continues for 12 rounds, and takes approximately 15 minutes.
    The final results for the N8 templates can be seen in Figure \ref{fig:N8_trained}.
    The templates are now a much better match to the photometry and more closely resemble physical galaxy spectra.
    Most of the templates have a Balmer Break at 4000 \AA, although this was essentially already present in the initial templates.
    In addition, there are now emission and absorption lines visible in the spectra at a much higher resolution than the broadband filters used for the photometry (some of which are labeled with gray lines in Figure \ref{fig:N8_trained}).
    Template N8-1 displays Mg and Na absorption lines and template N8-4 contains the beginnings of $H\alpha$ and $H\beta$ emission lines.
    Templates N8-6, N8-7, and N8-8 contain what appear to be H$\alpha$, H$\beta$, H$\gamma$, H$\delta$, OII, and OIII emission lines (see Section \ref{sect:speclines} for more analysis).
    The emergence of these high resolution features from a large ensemble of low resolution data is the one of the defining features of this method.

\begin{figure*}
        \centering
        \includegraphics{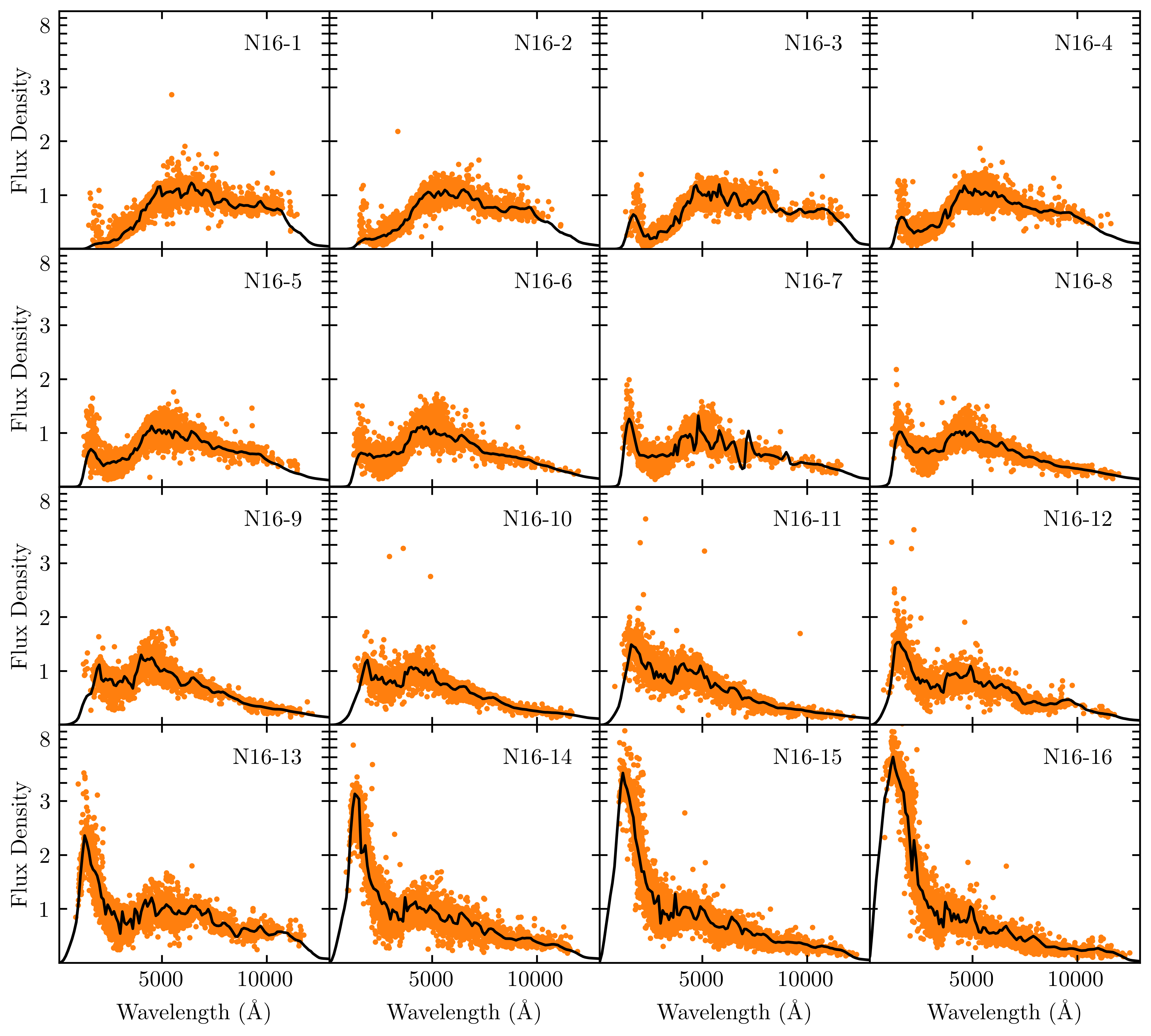}
        \caption{The trained N16 templates (black lines) with their final photometry sets (orange points). N16-1 is the reddest template, with each successive template getting bluer. These templates closely resemble the N8 templates and have show the same emerging spectral features (c.f. Figure \ref{fig:N8_trained}), but consist of a more continuous transition from red to blue spectra.}
        \label{fig:N16_trained}
    \end{figure*}

    In addition to these eight templates, we double the template number and train a set of 16 templates, in order to demonstrate the algorithm's ability to reconstruct templates with a more gradual transition of the colors from red to blue.
    This set (hereafter N16) was drawn from the same range of parameters for the log-normal function, and trained for 50 minutes over 26 rounds.
    The results of the training can be seen in Figure \ref{fig:N16_trained}.
    These results closely resemble the N8 results, with the same spectral features emerging.
    However, the N16 set shows a more gradual transition in color.

    In addition to starting from naive templates, one can start with templates derived from spectral synthesis models or observations of local galaxy spectra \citep{Budavari2000b, Csabai2000}.
    Here we apply the training algorithm to a standard set of SED templates commonly used for photo-z estimation (e.g. \bpz, see Section \ref{sect:bpz}).
    This set (hereafter CWW+SB4) consists of four templates from \citet{Coleman1980a} and two starburst templates from \citet{Kinney1996a}, the latter of which were added to account for faint blue galaxies in the HDF-N. 
    These six templates were recalibrated by \citet{Benitez2004a} to correct for systematic differences between the observed and predicted galaxy colors in the HDF-N and other spectroscopic catalogs. 
    In addition to these six, CWW+SB4 contains two synthetic starburst templates from \citet{Bruzual2003b}, added by \citet{Coe2006a} to account for even bluer galaxies in the UDF.

\begin{figure*}
        \centering
        \includegraphics{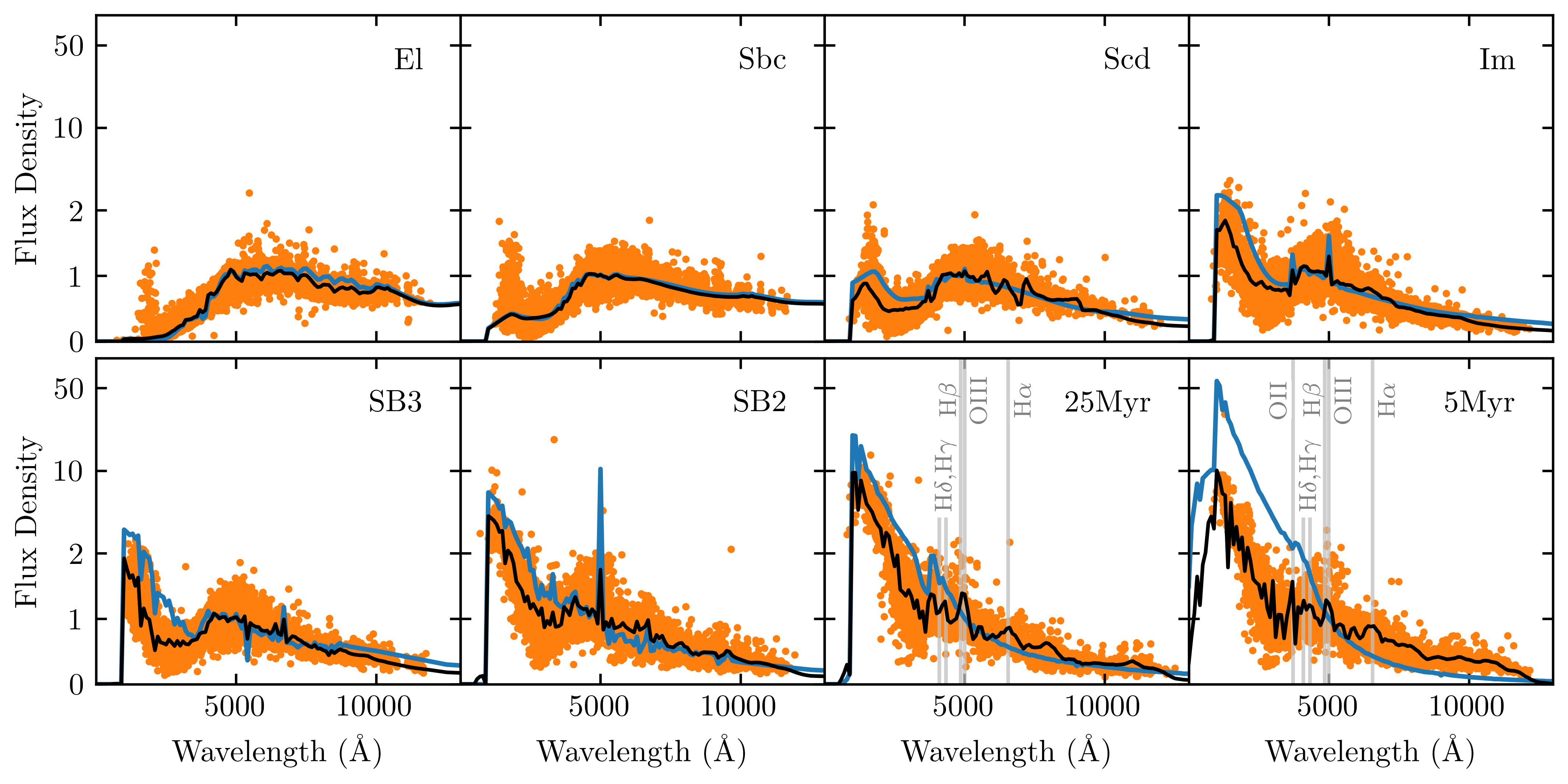}
        \caption{Result of training the CWW+SB4 templates. The original templates are in blue, the trained templates in black, and the final training sets are displayed as orange points. The 25Myr and 5Myr templates have acquired emission lines that were not present in the initial templates. These are labeled in gray.}
        \label{fig:cwwsb4_trained}
    \end{figure*}

    The CWW+SB4 templates were trained with $w=2$ for 46 minutes over 32 iterations.
    The results of the training can be seen in Figure \ref{fig:cwwsb4_trained}.
    The original templates are plotted in blue, with the trained templates plotted in black, along with the final photometry sets in orange.
    You can see that the El and Sbc templates have barely been altered. The remaining templates have all systematically become redder.
    The high resolution structure that was originally present in the Im, SB3, and SB2 templates have been decreased in magnitude, while additional structure has been added to the simulated 25Myr and 5Myr templates what were originally smooth.
    These new features have been labeled in gray.

    \subsection{Reconstructing Spectral Lines}
    \label{sect:speclines}

    The template training algorithm allows the reconstruction of high resolution spectral features from low resolution photometry due to the oversampling of the underlying SED templates.
    This includes the emergence of spectral lines in many of the templates (c.f. Figures \ref{fig:N8_trained}, \ref{fig:N16_trained}, and \ref{fig:cwwsb4_trained}).
    Knowledge of these lines allows us to perform post-processing of the learned templates to deconvolve the lines from the broadband filters.
    Here we perform a simple post-processing of the N8-6, N8-7, and N8-8 templates to reconstruct the emission lines labeled in Figure \ref{fig:N8_trained}.
    The templates are up-sampled to 10 \AA\ and the continuum of each is linearly interpolated around the emission lines.
    The excess flux is attributed to the corresponding spectral lines. 
    The flux of the H$\beta$ line is impossible to distinguish from the OIII line in our templates because they are so close to one another.
    The same is true for the H$\gamma$ and H$\delta$ lines.
    To overcome this difficulty, we use the Balmer decrements of $10^4$K SDSS galaxies from \citet{Groves2012a}: H$\alpha/$H$\beta = 2.86$ and H$\gamma/$H$\delta = 1.81$.
    We calculate the H$\beta$ flux from H$\alpha$, and subtract this from the combined H$\beta$-OIII flux, and we calculate H$\gamma$ and H$\delta$ from the combined H$\gamma$-H$\delta$ flux.

\begin{figure*}
        \centering
        \includegraphics{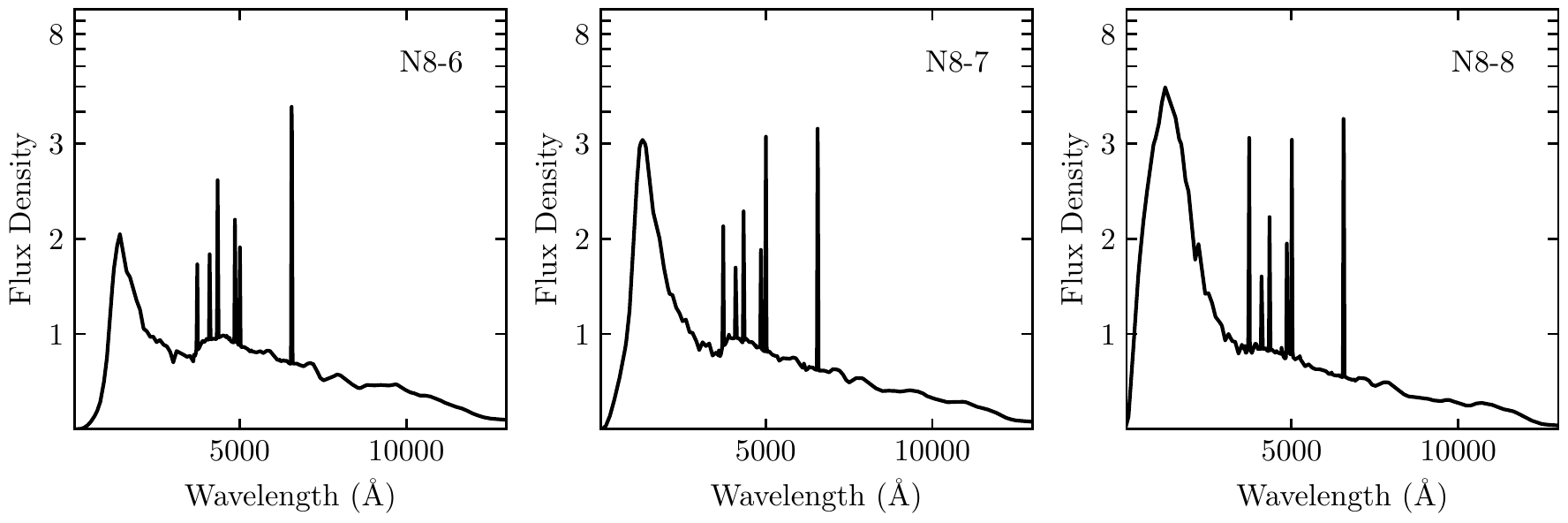}
        \caption{The N8-6, N8-7, and N8-8 templates with reconstructed emission lines (cf. Figure \ref{fig:N8_trained}). The emission lines, left to right, are OII,H$\delta$, H$\gamma$, H$\beta$, OIII, and H$\alpha$. The wavelengths, relative amplitudes, and effective widths of these lines are in Table \ref{tab:speclines}.}
        \label{fig:speclines}
    \end{figure*}

    After calculating the flux of the emission lines, the final templates are built by adding Gaussians of equivalent amplitude and FWHM = 20 \AA\ to the continuum. 
    The templates with the reconstructed spectral lines can be seen in Figure \ref{fig:speclines}.
    For each line, we calculate the amplitude relative to H$\beta$, and the effective width, $W_\lambda = \int (1 - F_\lambda/F_0) d\lambda$, where $F_\lambda$ is the total flux, and $F_0$ is the continuum flux.
    These values can be seen in Table \ref{tab:speclines}.
    Note that the amplitudes of our reconstructed H$\gamma$ and H$\delta$ lines relative to H$\beta$ are approximately three times greater than those listed in \citet{Groves2012a}.

    \begin{deluxetable}{c C C R C R C R }
        \tablecaption{Reconstructed Emission Lines \label{tab:speclines}}
        \tablehead{ \colhead{} & \colhead{} & \multicolumn{2}{c}{N8-6} & \multicolumn{2}{c}{N8-7} & \multicolumn{2}{c}{N8-8} \\ \cmidrule(lr){3-4} \cmidrule(lr){5-6} \cmidrule(lr){7-8} \colhead{Line} & \colhead{$\lambda$} & \colhead{$r$} & \colhead{$W_\lambda$} & \colhead{$r$} & \colhead{$W_\lambda$} & \colhead{$r$} & \colhead{$W_\lambda$}}
        \startdata
            H$\alpha$ & 6563 & 2.86 & 132.7 & 2.86 & 103.3 & 2.86 & 115.2 \\
            H$\beta$  & 4861 & 1.00 &  32.9 & 1.00 &  26.4 & 1.00 &  30.3 \\
            H$\gamma$ & 4340 & 1.18 &  36.5 & 1.31 &  31.6 & 1.28 &  37.1 \\
            H$\delta$ & 4102 & 0.65 &  19.6 & 0.72 &  16.7 & 0.71 &  20.7 \\
            OII       & 3727 & 2.04 &  58.1 & 1.27 &  32.0 & 0.74 &  24.4 \\
            OIII      & 5007 & 2.08 &  68.0 & 2.42 &  66.1 & 0.86 &  27.3 \\
        \enddata
        \tablecomments{ For each emission line, $r$ is the amplitude relative to H$\beta$, and $W_\lambda$ is the effective width in angstroms.}
    \end{deluxetable}

\section{Estimating Photo-z's}

    \label{sect:photoz}

    We evaluate the results of our template training algorithm by using our learned templates to estimate photo-z's for the test set of galaxies using the software package \bpz\ \citep{Benitez2000a}, and comparing the results to the spec-z's and the photo-z's estimated using the original CWW+SB4 templates.
    The test set consists of 20,496 galaxies (20\% of the total set), with mean redshift $z_\text{mean} = 0.69$, max redshift $z_\text{max} = 3.61$, and magnitudes $13.8 < i < 25.7$.
    See Table \ref{tab:data_sets} for a full summary and Figure \ref{fig:redshift_dist} for the redshift distribution.

    \subsection{Bayesian Photometric Redshifts}
    \label{sect:bpz}

    Bayesian Photometric Redshifts (\bpz; \citealt{Benitez2000a}) is a template-based photo-z estimator.
    Template-based estimators take a set of SED templates, assumed to be spanning and exclusive, and calculate observed fluxes over a grid of redshift values.
    For each template, \bpz\ evaluates a $\chi^2$ function at each redshift on the grid:
    \begin{align}
        \chi^2 (z,T,A) = \sum_n \frac{1}{\sigma_n^2} (\, A \, \hat{f}_n(z,T) - f_n \,)^2,
        \label{eq:chi2}
    \end{align}
    where $T$ denotes the template, $z$ denotes the redshift, $A$ is a normalization, and $\hat{f}_n$, $f_n$, and $\sigma_n$ denote the calculated flux, the observed flux, and the fractional error as in Equation \ref{eq:cost_function}. 
    The sum over $n$ is a sum over the filters for the set of observed fluxes. 
    \bpz\ then evaluates the likelihood for producing the observed galaxy fluxes: $p(\{f_n\}|z,T) \propto \exp{(-\chi^2/2)}$. 
    The redshift posterior is then calculated by marginalizing over the set of templates:
    \begin{align}
        p(z|\{f_n\},m_0) &= \sum_T \, p(z,T|\{f_n\},m_0) \nonumber \\
                        &\propto \sum_T \, p(z,T|m_0) \, p(\{f_n\}|z,T),
    \end{align}
    where $p(z,T|m_0)$ is a prior over the apparent magnitude $m_0$. 
    Work is underway to determine how best to use the full information encoded in the redshift posterior generated by \bpz\ and other photo-z codes (e.g. \citealt{Schmidt2020}). 
    In this work, however, only the mode of the posterior distribution is used to estimate the photo-z.

    We use \bpz-v1.99.3\footnote{\url{http://www.stsci.edu/~dcoe/BPZ/}} to estimate photo-z's.
    We turn off template interpolation by setting \texttt{INTERP=0}.
    For simplicity, we treat non-detections as non-observations.
    We use the various sets of SED templates described in Section \ref{sect:application}, and use the prior described in the following section.
    All other settings were left as default.

    \subsection{Galaxy Magnitude Priors}

    Before estimating photo-z's with \bpz, we must first construct the magnitude priors, $p(z,T|m_0)$, calibrated to the galaxies in our training set.
    We separate the prior into two parts:
    \begin{align}
        p(z,T|m_0) = p(T|m_0) \, p(z|T,m_0)
    \end{align}
    For the magnitude $m_0$, we use one of the $i$ bands in the following order of priority: $i$, $i_2$, $I$, $i^+$.
    Instead of constructing a different prior for each template, we follow \citet{Benitez2000a} in dividing our templates into three broad classifications: elliptical (El), spiral (Sp), or irregular/starburst (Im/SB).
    The CWW+SB4 templates are already classified under this scheme.
    We classify our new templates and each of the galaxies in the training set by assigning the classification of the CWW+SB4 template with the most similar colors, determined by minimizing the mean square error of the fluxes.
    
\begin{figure}
        \centering
        \includegraphics{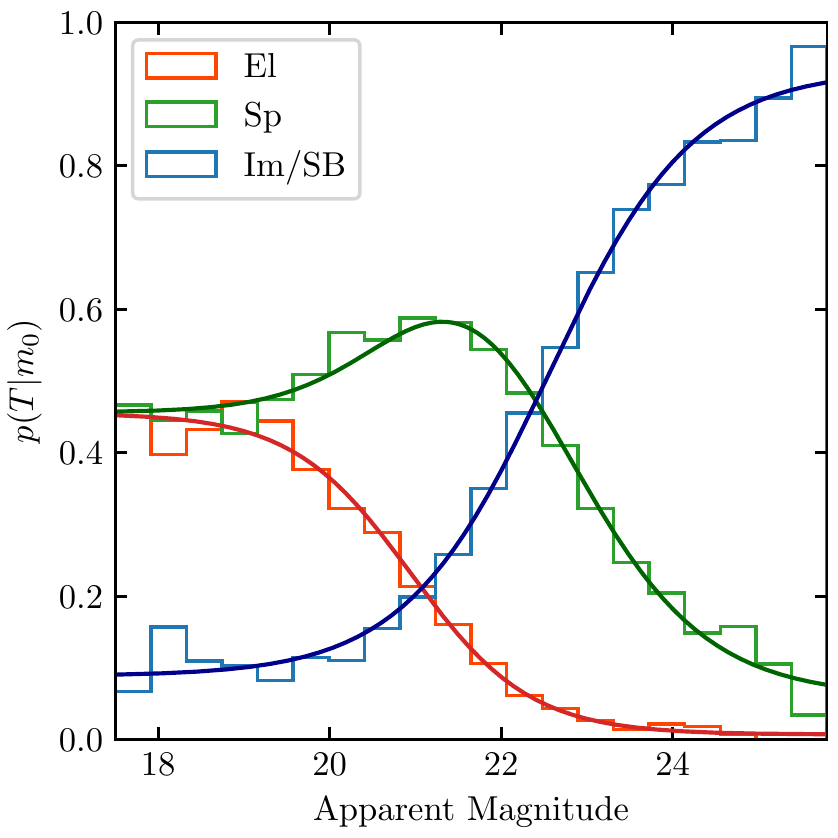}
        \caption{Fraction of each spectral class as a function of apparent magnitude. The histograms represent the fractions in the training set, and the curves are the spectral type priors fit to the data.}
        \label{fig:class_vs_mag}
    \end{figure}

    The N8 templates are determined to have one elliptical, four spiral, and three irregular/starburst galaxies; the N16 templates are determined to have two elliptical, eight spiral, and six irregular/starburst galaxies.
    The fraction of each classification as a function of magnitude for the training set galaxies is displayed in Figure \ref{fig:class_vs_mag}.

    We assume that the El and Im/SB galaxies have spectral priors of the form
    \begin{align}
        p(T|m_0) = \frac{L_T}{1+e^{-\kappa_T(m_0 - m_T)}} + C_T,
    \end{align}
    while $p(\text{Sp}|m_0) = 1 - p(\text{El}|m_0) - p(\text{Im/SB}|m_0)$.
    The values of $\{L_T,\kappa_T,m_T,C_T\}$ for the El and Im/SB galaxies are found by fitting to the distributions in Figure \ref{fig:class_vs_mag}.
    All three priors are plotted in the same figure, and the parameter values are listed in Table \ref{tab:prior_params}.

    For the redshift prior, we use Equations 23 and 24 from \citet{Benitez2000a}:
    \begin{align}
        p(z|T,m_0) = \frac{1}{N_T} \exp \left\{ -\left( \frac{z}{Z_T} \right)^{\alpha_T} \right\},
        \label{eq:z_prior}
    \end{align}
    where the normalization is
    \begin{align}
        N_T = \frac{Z_T^{~\alpha_T + 1}}{\alpha_T} ~ \Gamma \left( \frac{\alpha_T + 1}{\alpha_T} \right),
    \end{align}
    and the ``median'' redshift $Z_T$ is chosen to have the linear dependence
    \begin{align}
        Z_T(m_0) = z_{0T} + k_T(m_0 - 20).
    \end{align}
    Equation \ref{eq:z_prior} reproduces the exponential cutoff at high redshifts present in the training set, and can reasonably approximate any unimodal redshift distribution, from very narrow ($\alpha \gg 2$) to very broad ($\alpha \ll 1$).
    This flexibility reduces the bias introduced by the functional form of the prior \citep{Benitez2000a}.
    The nine parameters $\{\alpha_T, z_{0T}, k_T\}$ are determined by maximizing the likelihood $L = \prod_i p(z_i|T_i,m_{0i})$, where the product is over the galaxies in the training set.
    The parameters and their bootstrapped uncertainties are listed in Table \ref{tab:prior_params}.

    \begin{deluxetable*}{l C C C C C C C }
        \tablecaption{Parameters for the Priors, $p(z,T|m_0)$ \label{tab:prior_params}}
        \tablehead{\colhead{Spectral Type} & \colhead{$L_T$} & \colhead{$\kappa_T$} & \colhead{$m_T$} & \colhead{$C_T$} & \colhead{$\alpha_T$} & \colhead{$z_{0T}$} & \colhead{$k_T$}}
        \startdata
            El & 0.448 \pm 0.017 & -1.45 \pm 0.16 & 21.0 \pm 0.1 & 0.007 \pm 0.009 & 3.88 \pm 0.04 & 0.484 \pm 0.003 & 0.119 \pm 0.002 \\
            Sp & \ldots & \ldots & \ldots & \ldots & 3.40 \pm 0.04 & 0.493 \pm 0.003 & 0.124 \pm 0.002 \\
            Im/SB & 0.845 \pm 0.031 & \ \ 1.20 \pm 0.11 & 22.6 \pm 0.1 & 0.089 \pm 0.013 & 2.22 \pm 0.03 & 0.361 \pm 0.009 & 0.130 \pm 0.008 \\
        \enddata
    \end{deluxetable*}

    \subsection{Photo-z Results}
    \label{sect:photoz_results}

    We estimate photo-z's for the test set galaxies using \bpz\ with the settings and priors described in the previous two sections.
    We used four template sets: the original CWW+SB4 templates, the trained CWW+SB4 templates, and the trained N8 and N16 templates.

    \bpz\ provides two metrics for the photo-z estimates: \texttt{ODDS} and $\chi_{\text{mod}}^2$.
    \texttt{ODDS} measures how narrowly peaked the posterior distribution $p(z|\{f_n\},m_0)$ is around the estimated photo-z.
    Galaxies with low \texttt{ODDS} have either broad redshift posteriors, or posteriors with multiple peaks.
    $\chi_{\text{mod}}^2$ measures how well the best fit template at the predicted redshift matches the observed fluxes. 
    For more about these metrics, see Section 4 of \citet{Benitez2000a} and Section 4.3 of \citet{Coe2006a}.
    In this work, photo-z estimates with \texttt{ODDS} $< 0.95$ or $\chi_{\text{mod}}^2 > 1$ are excluded from the analysis, and the fraction excluded on this bases is reported as $f_\text{cut}$.

    To further evaluate the results of \bpz, we calculate the scatter, bias, and outlier fraction of the photo-z estimates. 
    Photo-z estimates are known to be contaminated with a significant number of outliers.
    This is largely driven by a degeneracy wherein the 1000\AA\ Lyman break in a high redshift galaxy spectrum has similar optical colors to the 4000\AA\ Balmer break in a low redshift galaxy spectrum. 
    \bpz\ attempts to break this degeneracy with the galaxy magnitude prior (i.e. galaxies with brighter apparent magnitudes are more likely to be at a lower redshift), yet there are still a large number of outliers.

    To address this issue, we evaluate the statistics of the interquartile range (IQR) of the data, as these measures are robust to the presence of outliers.
    We follow \citet{Graham2018a} in introducing the quantity $\Delta z_{1+z} = (z_{spec} - z_{phot})/(1 + z_{phot})$.
    The numerator quantifies the photo-z error and the denominator compensates for the larger uncertainty at high redshifts. 
    We define the scatter of the photo-z estimates, $\sigma_\text{IQR}$,  as the width of the IQR in $\Delta z_{1+z}$, divided by 1.349 to convert to the equivalent of a Gaussian standard deviation. 
    We define the bias of the photo-z estimates as the mean value of $\Delta z_{1+z}$ for galaxies within the IQR.
    The uncertainties of these two values are bootstrapped by calculating the values on 1000 random samples with replacement. 
    Outliers are identified as photo-z's with $\Delta z_{1+z} > 3 \sigma_{\text{IQR}}$, and the fraction of outliers is reported as $f_\text{out}$.

    \begin{figure*}
        \centering
        \includegraphics{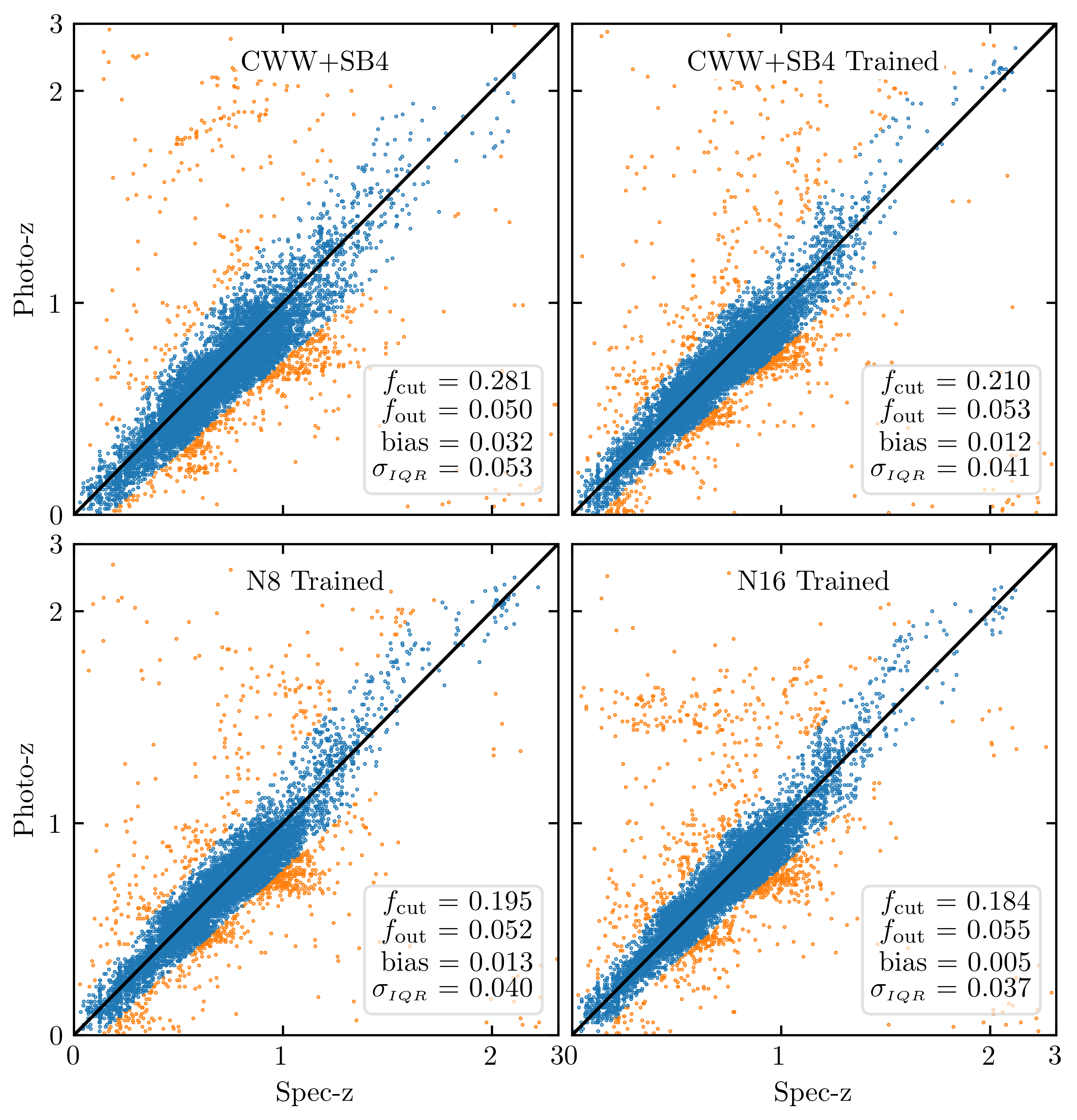}
        \caption{Results of photo-z estimation with \bpz, using the four different templates sets. Photo-z estimates are displayed as points: inliers are blue and outliers are orange. The black line represents perfect estimation (i.e. photo-z = spec-z). The statistics printed in each panel are for the entire data set.}
        \label{fig:photoz_results}
    \end{figure*}

    The photo-z results can be seen in Figure \ref{fig:photoz_results}.
    The photo-z estimates that passed the cuts on \texttt{ODDS} and $\chi_\text{mod}^2$ are displayed as points: the inliers in blue, the outliers in orange.
    The values of the photo-z statistics for each template set are printed in each panel.
    For all four template sets, the photo-z estimation is reasonably accurate for spec-z's $z < 1.5$.
    For higher redshifts, there appears to be a systematic bias towards higher photo-z's.
    Reduced photo-z accuracy is generally expected for spec-z's greater than 1.5, as the Balmer break leaves the optical bands at around $z=1.4$ and the Lyman break does not enter the ultraviolet bands until $z=2.5$.

    For the CWW+SB4 templates, the training algorithm decreased the fraction of photo-z's cut by 25\%, the bias by 63\%, and the scatter by 23\%, but did not improve the outlier fraction.
    We were able to achieve similar photo-z results using the trained N8 and N16 template sets, demonstrating that our training algorithm can be used to generate photo-z templates without any a priori information about galaxy spectra.
    Compared to the CWW+SB4 templates, N8 templates decreased $f_\text{cut}$ by 31\%, bias by 59\%, and scatter by 25\%.
    The N16 templates decreased $f_\text{cut}$ by 35\%, bias by 84\%, and scatter by 30\%.
    In all cases, the training algorithm decreases the fraction of bad photo-z's ($f_\text{cut} + f_\text{out}$), the bias, and the scatter.

\begin{figure*}
        \centering
        \includegraphics{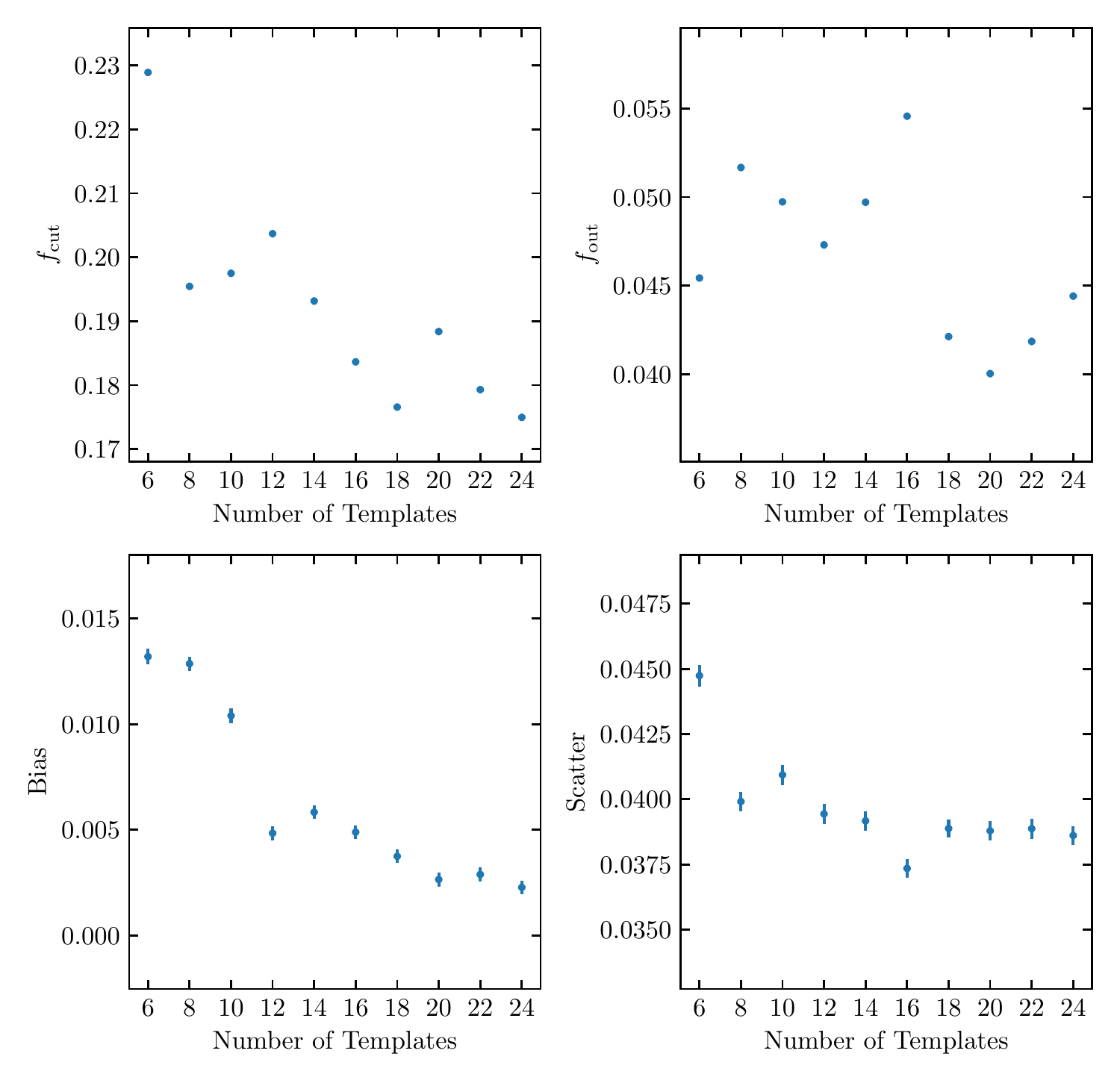}
        \caption{Photo-z statistics as a function of template number. Statistics are for the full redshift range.}
        \label{fig:Ntemplates}
    \end{figure*}

    Comparing the results for the N8 and N16 template sets indicate that increasing the number of templates can reduce the fraction cut, and the bias and scatter of the photo-z estimates.
    To further investigate this relationship, we calculate the photo-z statistics for a range of template numbers, the results of which are in Figure \ref{fig:Ntemplates}.
    We find that increasing the number of templates decreases the fraction cut and the bias, as well as slightly decreasing the scatter.
    The trend for outlier fraction is less clear.

    The N20 set has $f_\text{cut} = 0.188$ (a 33\% decrease compared to CWW+SB4), $f_\text{out} = 0.040$ (a 20\% decrease), bias = 0.003 (a 91\% decrease), and scatter = 0.039 (a 26\% decrease).

\begin{figure*}
        \centering
        \includegraphics{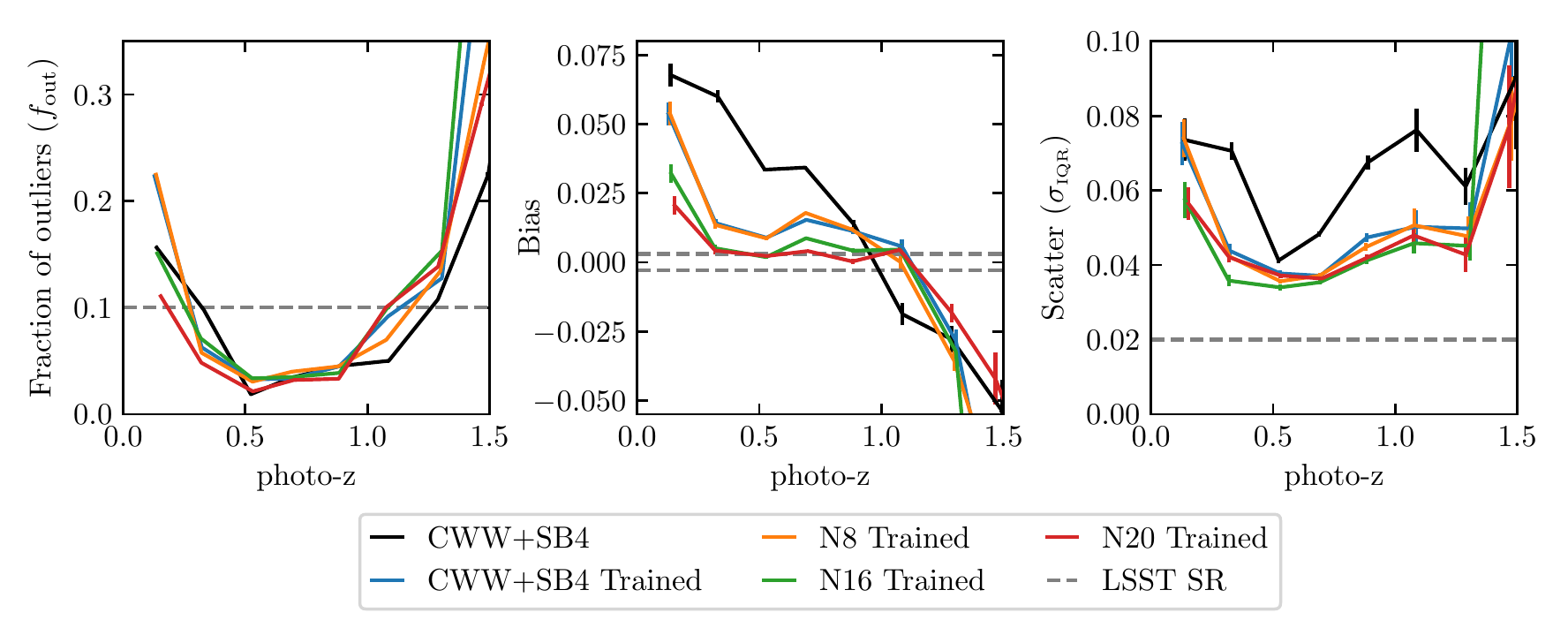}
        \caption{Photo-z metrics for the various template sets as a function of redshift bin. LSST science requirements are shown as dashed gray lines.}
        \label{fig:photoz_binned}
    \end{figure*}

    The value of the metrics as a function of photo-z can be seen in Figure \ref{fig:photoz_binned}.
    In addition to the template sets plotted above, we add the N20 set.
    For comparison, plotted in gray are the LSST science requirements for the metrics as listed in the LSST Science Requirement Document (SRD; \citealt{Ivezic2018}).
    The SRD lists the following minimum requirements to enable the envisioned LSST cosmological studies: root-mean-square error $< 0.02(1+z_\text{phot})$; $f_\text{out} < 10\%$; average bias $<0.003(1+z_\text{phot})$.
    The SRD lists these requirements for an $i<25$, magnitude-limited sample of four billion galaxies from $0.3 < z < 3.0$.
    For comparison, our test set consists of 20,496 galaxies with $i < 25.7$, in the range $z < 3.6$, including 19,391 galaxies with $i < 25$, in the range $0.3 < z < 3.0$.
    In Figure \ref{fig:photoz_binned}, we show that for redshifts $0.3 < z < 1.2$ we are able to achieve an appropriate outlier fraction, and that our training algorithm makes great progress on the bias, almost reaching the threshold required for LSST.
    We also make modest progress on the scatter, but reduction by another factor of two is still required.
    Beyond redshift $z=1.2$, all of our metrics fail the LSST science requirements.

\section{Discussion}

    \label{sect:discussion}

    In Section \ref{sect:template_training}, we demonstrated that our training algorithm could learn galaxy SED templates from photometry at a high resolution relative to the filters used to make the observations.
    We are able to learn a set of templates over twice the size of the standard CWW+SB4, showing a smooth progression of galaxy colors from red to blue.
    The spectra contain relatively high resolution spectral features, and post-processing can further reconstruct emission and absorption lines.
    The bluer templates contain more structure as they represent star forming galaxies and thus have stronger emission lines.
    In addition, the bluer templates have a larger number of high-redshift galaxies compared to the red templates, which aids the reconstruction of high-resolution features.
    While the high-redshift galaxies number in the hundreds instead of thousands, our results indicate that high-resolution features can be reliably reconstructed with only a few hundred galaxies.

    Our method has a number of limitations.
    The success of our algorithm relies on the ability to generate a naive set of templates as a starting point that will reliably divide the photometry by the spectral type of the galaxy. 
    This is relatively easy to accomplish for fewer than 20 templates, as was demonstrated by our simple photometry matching procedure and the log-normal templates we used.
    This is a strength of the algorithm as it is relatively robust to the starting templates. If, however, you wish to derive more than 20 templates from the photometry, care must be taken in the division of the photometry set to ensure there are sufficient galaxies in each subset to fully sample the entire wavelength range for the templates.
    In addition, the inherently discretized way in which we divide the photometry set stands in the way of generating a truly continuous set of SED templates.
    For a more continuous set of templates one might imagine taking two ``adjacent'' photometry sets, and assembling a photometry set ``between'' them by taking the bluer half on one set together with the redder half of the other. 
    Equally we could construct a moving window that progressively subdivides a sample based on color (with galaxies allowed to be present in more than one subet).

    Our data consists only of broadband photometry, however our algorithm would work equally well with narrow bands as well.
    Combining broadband and narrow band photometry would expand the data set and further constrain the templates.
    In particular, the addition of narrowband photometry should increase the resolution of spectral features recovered, and may allow one to resolve features such as the H$\gamma$ and H$\delta$ emission lines that we had to treat as a single feature.
    One could also include bands from a wider range of wavelengths to increase the wavelength range over which the templates are constrained.
    We attempted to include fluxes from the $K$-bands included with the zCOSMOS and VIPERS catalogs to learn  infra-red wavelengths for the templates, but there appeared to be systematic calibration issues in the data that we could not resolve.
    There is evidence that the inclusion of near-infrared and near-ultraviolet photometry in photo-z estimation can reduce outliers and scatter by up to 50\% each \citep{Graham2020}.

    In addition, for the results presented here, we used only galaxy fluxes with SNR greater than 20.
    One can use galaxies with lower SNR if outlier fluxes are removed from the photometry sets before training (we had success using an Isolation Forest; \citealt{Ting2008,Liu2012}).
    However, lowering the SNR of the photometry generally reduces the resolution of the structure you can reconstruct.

    The training algorithm itself could be made more sophisticated by restoring the wavelength dependence of the hyperparameter $\Delta_k$.
    We also hope to move beyond an iterative regression approach into deep learning, perhaps using Generative Adversarial Networks (GANs; \citealt{Goodfellow2014}).

    When constructing the \bpz\ prior, we sorted our templates into broad spectral classes.
    In the N8 set, for example, we determined that one template was elliptical, four were spiral, and three were irregular/starburst.
    Each of our templates has approximately the same number of galaxies matched to it, and the photometry matched to the elliptical templates does not display more variance than the photometry matched to other templates.
    These observations indicate that our data set contains a larger number of spiral and irregular/starburst galaxies than elliptical galaxies, rather than suggesting that the space of elliptical galaxy spectra is less finely sampled.
    For this reason, we do not expect the imbalance of the template number in each class to have a large impact on the photo-z quality, but nevertheless note that a more sophisticated prior could be constructed without relying on this broad classification scheme which may provide better redshift estimates.

    We found in Section \ref{sect:photoz_results} that our training algorithm can improve the bias and scatter of photo-z estimates.
    We found that increasing the number of templates enhances these improvements, with the best results for 20 templates.
    As mentioned above, with our current method for generating photometry sets, we struggle to reliably reconstruct more than 20 templates, so whether these benefits continue to decrease with template number is unknown.

    We can compare our method for generating more SED templates with \bpz's method of linearly interpolating between templates.
    N8 with \texttt{INTERP=2} generates 22 total templates.
    Table \ref{tab:interp_comparison} compares the photo-z results using these templates with the results using 22 templates learned from the photometry with \texttt{INTERP=0}.
    It is clear that, as far as $f_\text{out}$ and bias, our method for generating extra templates is superior to the linear interpolation used by \bpz. 

    \begin{deluxetable}{c C C C C C C}
        \tablecaption{Comparison to BPZ Interpolation \label{tab:interp_comparison}}
        \tablehead{ & \colhead{\texttt{INTERP}} & \colhead{Total N} & \colhead{$f_\text{cut}$} & \colhead{$f_\text{out}$} & \colhead{Bias} & \colhead{Scatter}}
        \startdata
            N8  & 0 &  8 & 0.228 & 0.058 & 0.014 & 0.040 \\
            N8  & 2 & 22 & 0.209 & 0.060 & 0.012 & 0.037 \\
            N22 & 0 & 22 & 0.214 & 0.045 & 0.004 & 0.039 \\
        \enddata
        \tablecomments{ Total N is the total number of SED templates in the set, including those interpolated by \bpz. Statistics quoted are for the full redshift range.}
    \end{deluxetable}

    The photo-z estimation with our learned template sets outperforms the results of the standard CWW+SB4 templates, however, more work needs to be done to reach the requirements set for LSST, especially for redshifts $z>1$.
    Templates can be trained for LSST science using the substantial overlap of LSST photometry with the eBoss \citep{Dawson2016} and Dark Energy Spectroscopic Instrument (DESI; \citealt{DESICollaboration2016}) surveys which will provide hundreds of thousands of spec-z's for LSST photo-z training and calibration \citep{Schmidt2014,Newman2015}.

    Our training method can be extended to other domains (e.g.\ stellar spectral reconstruction) where you can take a large set of incomplete data, segment that data into classes, and treat the set of unique observations in each class as an ensemble of observations of some class archetype, and thereby reconstruct more complete information.
    We plan to adapt the method to reconstruct supernova lightcurves from supernova photometry.
    
\section{Conclusions}
    \label{sect:conclusion}

    We have shown that galaxy SED templates can be learned directly from a data set of broadband photometry.
    Large sets of photometry at various redshifts can be leveraged to reconstruct high resolution features, such as the H$\alpha$, H$\beta$, H$\gamma$, H$\delta$, OII, and OIII emission lines, as well as Na and Mg absorption lines.
    Simple post processing can further improve the resolution of these reconstructed lines.
    The number of templates learned is variable and can be increased to more continuously sample the space of galaxy spectra and to improve photo-z results.

    We used our templates to estimate photo-z's for a test set of galaxies using \bpz.
    We found that training the standard set of templates that comes with \bpz\ decreases the fraction of bad photo-z's by 21\%, the bias by 63\% and the scatter by 23\%.
    Our own trained naive templates yielded better results.
    We learned a set of 20 templates from the data that reduced the fraction of bad photo-z's by 31\%, the bias by 91\%, and the scatter by 26\%.
    These derived templates outperform the interpolated spectra used by BPZ.
    The improvements in bias are almost sufficient to meet the requirements set for LSST, but another reduction by a factor of two is needed for the scatter.

    The templates derived with our training algorithm demonstrate that accurate galaxy spectra can be learned from broadband photometry.
    Our SED's could potentially be used for applications other than photo-z's, and our learning algorithm can be extended to other applications, such as learning supernova lightcurves from photometry.

    Our derived templates and the code used to produce these results are publicly available in a dedicated Github repository: \url{https://github.com/dirac-institute/photoz_template_learning}.

\acknowledgments

The authors would like to thank Bryce Kalmbach for providing advice in early stages of this work, Sam Schmidt for his help with \bpz, Melissa Graham for her code to calculate photo-z statistics, and Tam\'as Budav\'ari for comments on the manuscript. This work was supported by the U.S. Department of Energy, Office of Science, under Award DE-SC-0011635.

This research is based on observations made with ESO Telescopes at the La Silla or Paranal Observatories under programme ID(s) 175.A-0839(B), 175.A-0839(D), 175.A-0839(I), 175.A-0839(J), 175.A-0839(H), 175.A-0839(F).
This research is also based on observations obtained with MegaPrime/MegaCam, a joint project of CFHT and CEA/DAPNIA, at the Canada-France-Hawaii Telescope (CFHT) which is operated by the National Research Council (NRC) of Canada, the Institut National des Science de l'Univers of the Centre National de la Recherche Scientifique (CNRS) of France, and the University of Hawaii.
This research is also based in part on data products produced at Terapix available at the Canadian Astronomy Data Centre as part of the Canada-France-Hawaii Telescope Legacy Survey, a collaborative project of NRC and CNRS.
We use data from the VIMOS VLT Deep Survey, obtained from the VVDS database operated by Cesam, Laboratoire d'Astrophysique de Marseille, France
We also use data from the VIMOS Public Extragalactic Redshift Survey (VIPERS).
VIPERS has been performed using the ESO Very Large Telescope, under the "Large Programme" 182.A-0886. 
The participating institutions and funding agencies are listed at \url{http://vipers.inaf.it}
This research has also made use of the SVO Filter Profile Service (\url{http://svo2.cab.inta-csic.es/theory/fps/}) supported from the Spanish MINECO through grant AYA2017-84089.

\software{Astropy \citep{AstropyCollaboration2013}, BPZ \citep{Benitez2000a}, Jupyter \citep{Kluyver2016}, Matplotlib \citep{Hunter2007}, Numpy \citep{VanDerWalt2011}, Scikit-learn \citep{Pedregosa2011}, Scipy \citep{Virtanen2020}.}
    
\newpage
\bibliography{references.bib}{}
\bibliographystyle{aasjournal}

\end{document}